\documentclass[aps,prd,preprint,endfloats,
               showpacs,superscriptaddress,nofootinbib]{revtex4}
\usepackage{graphicx}
\begin{document}
\setlength{\baselineskip}{16pt}
%
%
\title{
$I=2$ Pion Scattering Length from Two-Pion Wave Functions
}
%
%
\author{        S.~Aoki }
\affiliation{
Graduate School of Pure and Applied Sciences,
University of Tsukuba,
Tsukuba, Ibaraki 305-8571, Japan
}
\author{    M.~Fukugita }
\affiliation{
Institute for Cosmic Ray Research,
University of Tokyo,
Kashiwa 277 8582, Japan
}
\author{  K-I.~Ishikawa }
\affiliation{
Department of Physics,
Hiroshima University,
Higashi-Hiroshima, Hiroshima 739-8526, Japan
}
\author{    N.~Ishizuka }
\affiliation{
Graduate School of Pure and Applied Sciences,
University of Tsukuba,
Tsukuba, Ibaraki 305-8571, Japan
}
\affiliation{
Center for Computational Sciences,
University of Tsukuba,
Tsukuba, Ibaraki 305-8577, Japan
}
\author{     Y.~Iwasaki }
\affiliation{
Graduate School of Pure and Applied Sciences,
University of Tsukuba,
Tsukuba, Ibaraki 305-8571, Japan
}
\author{      K.~Kanaya }
\affiliation{
Graduate School of Pure and Applied Sciences,
University of Tsukuba,
Tsukuba, Ibaraki 305-8571, Japan
}
\author{      T.~Kaneko }
\affiliation{
High Energy Accelerator Research Organization (KEK),
Tsukuba, Ibaraki 305-0801, Japan
}
\author{   Y.~Kuramashi }
\affiliation{
Graduate School of Pure and Applied Sciences,
University of Tsukuba,
Tsukuba, Ibaraki 305-8571, Japan
}
\affiliation{
Center for Computational Sciences,
University of Tsukuba,
Tsukuba, Ibaraki 305-8577, Japan
}
\author{       M.~Okawa }
\affiliation{
Department of Physics,
Hiroshima University,
Higashi-Hiroshima, Hiroshima 739-8526, Japan
}
\author{       A.~Ukawa }
\affiliation{
Graduate School of Pure and Applied Sciences,
University of Tsukuba,
Tsukuba, Ibaraki 305-8571, Japan
}
\affiliation{
Center for Computational Sciences,
University of Tsukuba,
Tsukuba, Ibaraki 305-8577, Japan
}
\author{    T.~Yamazaki }
\altaffiliation[ Present address : ]{
RIKEN BNL Research Center,
Brookhaven National Laboratory,
Upton, NY 11973, USA
}
\affiliation{
Graduate School of Pure and Applied Sciences,
University of Tsukuba,
Tsukuba, Ibaraki 305-8571, Japan
}
\author{  T.~Yoshi\'{e} }
\affiliation{
Graduate School of Pure and Applied Sciences,
University of Tsukuba,
Tsukuba, Ibaraki 305-8571, Japan
}
\affiliation{
Center for Computational Sciences,
University of Tsukuba,
Tsukuba, Ibaraki 305-8577, Japan
}
\collaboration{ CP-PACS Collaboration }
%
%
\date{ \today }
%
%
\begin{abstract}
We calculate the two-pion wave function
in the ground state of the $I=2$ $S$-wave system
and find the interaction range between two pions,
which allows us to examine the validity of the necessary condition
for the finite-volume method for the scattering length
proposed by L\"uscher.
We work in the quenched approximation employing
a renormalization group improved gauge action for gluons
and an improved Wilson action for quarks
at $1/a=1.207(12)\ {\rm GeV}$
on
$16^3 \times 80$,
$20^3 \times 80$ and
$24^3 \times 80$ lattices.
We conclude that the necessary condition is satisfied
within the statistical errors
for the lattice sizes $L\ge 24$ ($3.92\ {\rm fm}$)
when the quark mass is in the range that corresponds to
$m_\pi^2 = 0.273-0.736\ {\rm GeV}^2$.
We obtain the scattering length with a smaller statistical error
from the wave function than from the two-pion time correlator.
\end{abstract}
\maketitle
%
%
\section{ Introduction }
\label{Sec:Introduction}
Calculations of the scattering length and the phase shift
represent an important step for expanding our understanding
of the strong interaction based on lattice QCD
to dynamical aspects of hadrons.
For the simplest case of the two-pion system in the $I=2$ $S$-wave system,
the scattering length has been calculated
in~\cite{SGK:a0,GPS:a0,KFMOU:a0,AJ:a0,Juge:a0,JLQCD:a0,LZCM:a0,DMML:a0,BGR:a0}
and the pioneering study of the phase shift
was made by Fiebig {\it et al.}~\cite{Fiebig:phsh}
using the two-pion effective potential. 
We presented a direct calculation of the phase shift
without recourse to the effective potential
in quenched~\cite{CP-PACS:phsh} and full QCD~\cite{CP-PACS:phsh_full}.
Kim reported on preliminary results of the phase shift
on $G$- and $H$-periodic boundary lattices~\cite{Kim:phsh}.

The calculation of the scattering length
and the phase shift usually employs the finite-volume method of L\"uscher,
in which the scattering phase shift is related to
the energy eigenvalue on a finite volume~\cite{Luscher_one:formula,
Luscher_LH:formula,LuscherWolf:formula,Luscher_two:formula}.
In previous application of the formula
the energy eigenvalues were
calculated from the asymptotic time behavior 
of the two-pion time correlator.

The derivation of L\"uscher's formula assumes the condition $R<L/2$
for the two-pion interaction range $R$ and the lattice size $L$,
so that the boundary condition does not distort
the shape of the two-pion interaction.
In the studies to date, however, calculations were carried out
without verifying this necessary condition,
but simply employing a few lattices
having different sizes and extrapolating to the infinite volume limit
assuming simple functions,
such as an inverse power of the lattice extent,
for finite volume corrections.
In the absence of theoretical justification, however,
such assumptions would cause ambiguities and
it is important to examine the validity of the condition
in lattice simulations for reliable results.

In this work we restrict ourselves
to the ground state of the $I=2$ $S$-wave two-pion system
and calculate the two-pion wave function.
We investigate the two-pion interaction range
and the validity of the necessary condition for L\"uscher's formula.
We attempt to extract the scattering length directly from the wave function
and compare it with the more conventional result from
the two-pion time correlator.

We refer to the work for two-dimensional statistical models in~\cite{Wisz:stat}.
For the two-pion system of QCD a similar idea was discussed in~\cite{LMST:kpp}.
We also quote Yamazaki who presented preliminary results for the wave function
for the four-dimensional Ising model
and the scattering phase shift therefrom~\cite{Yamazalo:ISppwf}.

This paper is organized as follows.
In Sec.~\ref{Sec:Luscher}
we give a brief review of the derivation
of L\"uscher's formula~\cite{Luscher_two:formula}
with emphasis on the role of the condition $R<L/2$.
The calculational method of the wave function
and the simulation parameters
are given in Sec.~\ref{Sec:Method}.
In Sec.~\ref{Sec:Results_zan24}
we present the wave function and estimate
the two-pion interaction range for a $24^3$ lattice.
In Sec.~\ref{Sec:Results_SCL}
we calculate the scattering length from the wave function
and compare it with that from the two-pion time correlator.
Our investigations on $20^3$ and $16^3$ lattices
are given in Sec.~\ref{Sec:Results_zan_small},
where finite volume effects on the scattering length are examined
by comparing three lattice volumes.
Our conclusions are given in Sec.~\ref{Sec:Conclusions}.
Preliminary reports of the present work
were presented in~\cite{IY:ppwf}.
%
%
\section{ L\"uscher's formula }
\label{Sec:Luscher}
We briefly review the derivation of L\"uscher's formula,
with emphasis on the role of the condition for the two-pion interaction range.
The formula~\cite{Luscher_one:formula,Luscher_LH:formula}
was rederived using an effective Schr\"odinger equation
for the two-dimensional scalar filed theory in~\cite{LuscherWolf:formula},
and for the four-dimensional case in~\cite{Luscher_two:formula}
which is discussed here.
Another approach based on the Bethe-Salpeter wave function
in quantum field theory~\cite{LMST:kpp}
is discussed in Appendix.~\ref{appe:Luscher-RFT}.

The static two-pion wave function $\phi(\vec{x};k)$
with the energy eigenvalue $E=2\sqrt{k^2+m_\pi^2}$
in the center of mass system on a finite periodic box
of volume $L^3$ satisfies the effective Schr\"odinger
equation~\cite{Luscher_one:formula,LuscherWolf:formula} :
\begin{equation}
  ( \triangle + k^2 ) \phi(\vec{x};k)
  = \int {\rm d}^3 y \ U_k(\vec{x},\vec{y}) \phi(\vec{y};k)
\ ,
\label{Luscher_effEQ.eq}
\end{equation}
where $\vec{x}$ and $\vec{y}$
are the relative coordinate of the two pions.
$U_k(\vec{x},\vec{y})$ is
the Fourier transform of the modified Bethe-Salpeter kernel
for the two-pion interaction
on the finite volume~\cite{Luscher_one:formula},
and is related to the off-shell two-pion scattering amplitude
(see Appendix~\ref{appe:Luscher-RFT}).
It is generally non-local and depends on the two-pion energy.
It should be noticed that $k^2$ in (\ref{Luscher_effEQ.eq})
is not a square of a 3-dimensional momentum vector
but defined from the energy by $k^2 = E^2/4 - m_\pi^2$.
It may take a negative value in some cases.
We call $k$ ``momentum'' for simplicity in this paper, however.

In the derivation of L\"uscher's formula
it is assumed that
the two-pion interaction range is smaller than one half the lattice extent,
{\it i.e.} there exists the distance $R<L/2$,
where the wave function satisfies the Helmholtz equation :
\begin{equation}
  ( \triangle + k^2 ) \phi(\vec{x};k) = 0
     \quad \mbox{ for $\vec{x} \in V_R$ }
\ ,
\label{AsumeU.eq}
\end{equation}
with
\begin{equation}
   V_R = \{ \ \vec{x} \ | \ | \vec{x} + \vec{n} L | > R
                    \ , \ \vec{n} \in {\bf Z}^3 \ \}
\ .
\label{def_VR.eq}
\end{equation}

Next we consider solutions of (\ref{AsumeU.eq}).
In general $k^2 L^2 / (2\pi)^2$
can be integer or non-integer.
The former case is called ``singular-value solutions''
in~\cite{Luscher_two:formula} (see also~\cite{LMST:kpp}).
The appearance of these solutions is not generic.
It occurs only for some specific lattice volumes
or particular cases of the two-pion interaction.
For the two-pion ground state there is
an important singular-value solution that has $k^2=0$,
which, however, exists only on specific lattice volumes
or for the vanishing scattering length.
If this solution appears in numerical simulations,
the two-pion time correlator should behave as
\begin{equation}
  \frac{ \langle 0 | \pi(t)\pi(t) \pi(0)\pi(0) | 0 \rangle }
       { \langle 0 | \pi(t) \pi(0) | 0 \rangle^2 }
    \sim \mbox{ const.}
\ ,
\label{sing_sol.eq}
\end{equation}
for a large $t$.
However, such a time behavior has not been observed in previous studies
or in our numerical simulations.
Thus such a case hardly occurs for the ground state.
The formula for the singular-value solutions
was derived in Ref.~\cite{Luscher_two:formula},
but we consider only non-integer value solutions in this paper.

General solutions of the Helmholtz equation (\ref{AsumeU.eq})
can be written by
\begin{equation}
  \phi(\vec{x};k)
  =   \sum_{l=0}^{\infty}
      \sum_{m=-l}^{l}  \ v_{lm}(k) \cdot G_{lm}( \vec{x}; k )
\ ,
\label{Luscher_wave_one.eq}
\end{equation}
with $v_{lm}(k)$.
$G_{lm}(\vec{x};k)$ is given from
the periodic Green function :
\begin{equation}
  G(\vec{x};k)
  =   \frac{ 1 }{ L^3 }
      \sum_{\vec{p} \in \Gamma }
      \ \frac{1}{ p^2 - k^2 }
      \ {\rm e}^{ i \vec{p} \cdot \vec{x} }
\ ,
\quad
  \Gamma = \bigl\{ \ \vec{p} \ | \ \vec{p} = \vec{n} \frac{ 2\pi }{ L }
   \ , \ \vec{n} \in {\bf Z}^3 \bigr\}
\ ,
\label{Luscher_wave_Glm.eq}
\end{equation}
as
\begin{equation}
  G_{lm}(\vec{x};k) = \sqrt{4\pi} {\cal Y}_{lm}(\nabla) G(\vec{x};k)
\ ,
\label{Glm_def.eq}
\end{equation}
where ${\cal Y}_{lm}(\vec{x})$ is a polynomial
related to the spherical harmonics
through ${\cal Y}_{lm}(\vec{x}) = x^l \cdot Y_{lm}(\Omega_x)$
with $\Omega_x$ the spherical coordinate for $\vec{x}$
and $x=|\vec{x}|$.
The convention of $Y_{lm}(\Omega)$ is that of~\cite{MESSIAH:book},
as is adopted in~\cite{Luscher_two:formula}.
It then follows that
$G_{00}(\vec{x};k) = G(\vec{x};k)$.

In general we can expand the solution of the Helmholtz equation
in terms of the spherical Bessel $j_l(kx)$ and Neumann functions $n_l(kx)$
for $R<x<L/2$ as
\begin{equation}
  \phi( \vec{x}; k ) =
    \sum_{l=0}^{\infty}
    \sum_{m=-l}^{l}
      \sqrt{4\pi} Y_{lm}(\Omega_x)
        \biggl(      \alpha_l(k) \cdot j_l(kx)
                  +  \beta_l (k) \cdot n_l(kx)
        \biggr)
\ ,
\label{Luscher_wave_exp.eq}
\end{equation}
where the conventions of $j_l(x)$ and $n_l(x)$
agree with those in~\cite{MESSIAH:book} and~\cite{Luscher_two:formula}.
The expansion coefficients $\alpha_l(k)$ and $\beta_l(k)$
yield the scattering phase shift in the infinite volume as
$\tan\delta_l(k) = \beta_l(k) / \alpha_l(k)$.
In particular, for the ground state
the momentum $k^2$ is very small
and the $S$-wave scattering length $a_0$ is given by
$\tan\delta_0(k) = \beta_0(k) / \alpha_0(k) = a_0 k + {\rm O}(k^3)$.

For the wave function (\ref{Luscher_wave_one.eq}),
$\beta_l (k)$ and $\alpha_l (k)$ are geometrically related,
because they can be expressed in terms of the expansion coefficients
for $G_{lm}(\vec{x};k)$.
The expansion of $G(\vec{x};k)$ is given by
\begin{equation}
  G (\vec{x};k)
  = \frac{k}{4\pi} n_0(kx)
    +
    \sum_{l=0}^{\infty}
    \sum_{m=-l}^{m=l}
           \sqrt{4\pi} Y_{lm}(\Omega_x)
           g_{lm}(k) j_l(kx)
\ ,
\label{Exp_G.eq}
\end{equation}
where
\begin{equation}
  g_{lm}(k)
  =
    \frac{1}{L^3}
    \sum_{\vec{p} \in \Gamma}
          \frac{ ( i p / k )^l }{ p^2 - k^2 }
          \sqrt{4\pi} Y_{lm}(\Omega_p)
\ ,
\quad
  \Gamma = \bigl\{ \ \vec{p} \ | \ \vec{p} = \vec{n} \frac{ 2\pi }{ L }
   \ , \ \vec{n} \in {\bf Z}^3 \bigr\}
\label{Sperical_zeta.eq}
\end{equation}
with spherical coordinate $\Omega_p$ for $\vec{p}$.
(The function $g_{lm}(k)$
differs from $g_{lm}$ in (3.31) of Ref.~\cite{Luscher_two:formula}
only by the normalization as
$g_{lm}(k)=1/\sqrt{4\pi}\cdot g_{lm}$).
The explicit expansion for $G_{lm}(\vec{x};k)$ with general $l$ and $m$
is not needed.
Note that the indices $l$ and $m$
are not labels of the angular momentum,
nor $G_{lm}(\vec{x};k)$ is the eigenstate of the angular momentum
labeled by $l$ and $m$.
Actually it includes $Y_{l'm'}(\Omega_x)$ with $l'\not=l$ and $m'\not=m$.
Also note that
$G_{lm}(\vec{x};k)$ contains
$j_{l'}(kx)$ for a range of $l'$ and 
$n_{l'}(kx)$ with only $l'=l$.
These are easily known from (\ref{Glm_def.eq}) and (\ref{Exp_G.eq}).

In this work
we consider only wave functions in the ${\bf A_1^+}$ representation
of the cubic group, which equals $S$-wave up to angular momenta $l\ge4$.
The wave function for $\vec{x} \in V_R$ can be expressed as
\begin{equation}
  \phi(\vec{x};k) =
         v_{00}(k) \cdot G     (\vec{x};k)
      +  v_{40}(k) \cdot \frac{1}{48} \sum_{ {\bf R} } G_{40}({\bf R}[\vec{x}];k)
      +  \cdots
\ ,
\label{Luscher_wave_two.eq}
\end{equation}
where a vector operation ${\bf R}$ represents an element of the cubic group
which has 48 elements.
The terms with $l\ge 6$ are neglected,
and the $l=4$ terms with $m\ne 0$ do not appear
since they either vanish (for $|m|\not= 4$) or reduce to
$G_{40}(\vec{x};k)$ (for $|m|=4$).
If the scattering phase shift $\delta_l(k)$ with $l\ge 4$
is negligible in the energy range under consideration,
$\beta_l(k)\sim 0$ for $l\ge 4$ in (\ref{Luscher_wave_exp.eq}).
This means that
$v_{40}(k) \sim 0$ in (\ref{Luscher_wave_two.eq}),
and thus
\begin{equation}
  \phi(\vec{x};k)\sim v_{00}(k) \cdot G(\vec{x};k)
\ ,
\label{pppp.eq}
\end{equation}
because $G_{lm}(\vec{x};k)$ contains $n_l(kx)$.
This expectation is supported for the ground state
by our numerical simulations.

Finally we obtain L\"uscher's formula
between the $S$-wave scattering phase shift and the allowed value of $k^2$
by comparing the coefficients of $j_{0}(kx)$ and $n_{0}(kx)$
in (\ref{Exp_G.eq}),
\begin{equation}
  \frac{ 1 }{ \tan \delta_0( k )  }
  = \frac{ \alpha_0(k) }{ \beta_0(k) }
  = \frac{ g_{00}( k ) }{ k / (4\pi) }
  = \frac{4\pi}{k} \cdot
      \frac{1}{L^3}
      \sum_{\vec{p} \in \Gamma}
          \frac{ 1 }{ p^2 - k^2 }
\ ,
\label{Lusher.eq}
\end{equation}
where the region $\Gamma$ is defined by (\ref{Sperical_zeta.eq})
and (\ref{pppp.eq}) is assumed.

We remark that contaminations from inelastic scattering
are likely negligible for the ground state,
although they may become significant for momentum excitation states,
whose energies are close to or above the threshold
of inelastic scattering, $E \sim 4m_\pi$.
%
%
\section{ Method of calculations }
\label{Sec:Method}
%
%
\subsection{ Calculation of the wave function }
\label{SubSec:ppwf}
Our definition of the two-pion wave function is
\begin{equation}
   \phi(\vec{x};k) =
        \langle 0 | \ \Omega(\vec{x},t) \ | \pi^+ \pi^+ ; k  \rangle
        \cdot {\rm e}^{ 2 E_k t }
\ ,
\label{def_of_ppwf.eq}
\end{equation}
where $|\pi^+ \pi^+ ;k \rangle$ is the ground state
with energy $2 E_k=2 \sqrt{ m_\pi^2 + k^2 }$.
A factor $\exp ( 2E_k t )$ is introduced
to compensate the $t$ dependence.
The operator $\Omega(\vec{x},t)$
is an interpolating field for the two-pion,
which is defined by
\begin{equation}
   \Omega (\vec{x},t)
     = \frac{1}{48 } \sum_{ {\bf R} }
       \frac{1}{L^3} \sum_{ \vec{X} }
               \ \pi^+ ( {\bf R}[\vec{x}] + \vec{X},t)
               \ \pi^+ (                    \vec{X},t)
\ ,
\label{def_of_ppwf_op.eq}
\end{equation}
where
$\pi^+ (\vec{x},t) = \bar{u}(\vec{x},t) \gamma_5 d(\vec{x},t)$
is an interpolating operator for $\pi^+$ at $(\vec{x},t)$
and a vector operation ${\bf R}$ represents an element of the cubic group
which has 48 elements.
The summation over ${\bf R}$ and $\vec{X}$
projects out the ${\bf A_1^{+}}$ sector of the cubic group and
the zero total momentum state.
The wave function in (\ref{def_of_ppwf.eq})
is the Bethe-Salpeter wave function
projected to the ${\bf A_1^{+}}$ sector.
It has the
same properties as those of the wave function in Sec.~\ref{Sec:Luscher}
and we can derive L\"uscher's formula from it.
Details are discussed in Appendix~\ref{appe:Luscher-RFT}.
The wave functions at all positions $\vec{x}$ are not independent.
The number of independent position vectors
is $N(N+1)(N+2)/6$ with $N=L/2+1$
owing to the periodic boundary condition
$\phi(\vec{x};k) = \phi( \vec{x} + \vec{n}L ; k)$
with $\vec{n} \in {\bf Z}^3$
and the invariance under the cubic group
$\phi(\vec{x};k)=\phi({\bf R}[\vec{x}];k)$.

In order to calculate the wave function
we construct the correlator :
\begin{equation}
  F_{\pi\pi} (\vec{x},t;t_0,t_0+1) =
      \langle 0 | \ \Omega (\vec{x},t) \ W(t_0)\ W(t_0+1)\  | 0 \rangle
\ ,
\label{wave_ext.eq}
\end{equation}
where $W(t)$ is a wall source at time $t$ defined by
\begin{equation}
  W(t) = \frac{1}{L^3} \sum_{\vec{x}}
         \frac{1}{L^3} \sum_{\vec{y}}
               \bar{d}(\vec{x},t) \gamma_5 u(\vec{y},t)
\ ,
\label{wall_def.eq}
\end{equation}
which is used on configurations fixed to the Coulomb gauge.
The two wall sources are placed at different time slices $t_0$ and $t_0+1$
to avoid contaminations from Fierz-rearranged terms~\cite{KFMOU:a0}.
Neglecting contributions from the momentum excitation states
in the large $t$ region,
we obtain the wave function at the time slice $t$ as
\begin{equation}
  \phi(\vec{x};k)
  = \frac{ F_{\pi\pi}(\vec{x},t;t_0,t_0+1) }{ F_{\pi\pi}(\vec{x}_0,t;t_0,t_0+1) }
\ ,
\label{method_wfunc.eq}
\end{equation}
up to the overall constant,
where $\vec{x}_0$ is the reference position.

We try to extract the energy eigenvalue of the two-pion
and the scattering length from the wave function $\phi(\vec{x};k)$,
but, for comparison,
we also estimate them from the two-pion time correlator :
\begin{equation}
  F_{\pi\pi}(t;t_0,t_0+1)
    = \frac{1}{L^3} \sum_{\vec{x}} F_{\pi\pi}(\vec{x},t;t_0,t_0+1)
\ .
\label{def_pp_tfunc.eq}
\end{equation}
The single pion time correlator computed
with the aid of the wall source,
\begin{equation}
  F_\pi(t;t_0) = \frac{1}{L^3} \sum_{\vec{x}}
             \langle 0 | \pi^+(\vec{x},t) W(t_0) | 0 \rangle
\ ,
\label{def_p_tfunc.eq}
\end{equation}
is used to construct the normalized two-pion correlator :
\begin{equation}
  R(t) = \frac{ F_{\pi\pi}(t;t_0,t_0+1) }{ F_\pi(t;t_0) F_\pi(t,t_0+1) }
\ .
\label{R_t_def.eq}
\end{equation}
In the absence of the singular-value solution that belongs to $k^2=0$,
this behaves for a large $t$ as
\begin{equation}
  R(t) = C \cdot {\rm e}^{ - \Delta W \cdot ( t - t_0 ) }
\ ,
\label{R_t_dep.eq}
\end{equation}
where $C$ is a constant and
\begin{equation}
  \Delta W = 2E_k - 2m_\pi = 2\sqrt{ m_\pi^2 + k^2 } - 2m_\pi
\label{eq:deltaW}
\end{equation}
is the energy shift due to the two-pion interaction on the finite volume.
The momentum $k^2$ is calculated from
$\Delta W$, and it can be used to estimate the scattering length
via L\"uscher's formula.
%
%
\subsection{ Simulation parameters }
\label{Sec:Param}
Our simulation is carried out in quenched lattice QCD
employing a renormalization group improved gauge action for gluons,
\begin{equation}
  S_G = \frac{\beta}{6} \sum_x
        \biggl(
             C_0 \sum_{ \mu < \nu } W_{\mu\nu}^{1\times 1}(x)
           + C_1 \sum_{ \mu , \nu } W_{\mu\nu}^{1\times 2}(x)
        \biggr)
\ .
\label{RG_g.eq}
\end{equation}
The coefficient $C_1 = -0.331$
of the $1 \times 2$ Wilson loop $W_{\mu\nu}^{1\times 2}(x)$
is fixed by a renormalization group analysis~\cite{RG-gauge:IW},
and $C_0=1-8C_1=3.648$
of the $1 \times 1$ Wilson loop $W_{\mu\nu}^{1\times 1}(x)$
by the normalization condition,
which defines the bare coupling $\beta=6/g^2$.
Our calculation is carried out at $\beta=2.334$.
Gluon configurations are generated with the $5$-hit heat-bath algorithm
and the over-relaxation algorithm mixed in the ratio of $1:4$.
The combination is called a sweep
and physical quantities are measured every 200 sweeps.

We use an improved Wilson action for the quarks~\cite{SWW:SW}
with the clover coefficient $C_{SW}$ being the
mean-field improved choice defined by
\begin{equation}
  C_{SW}=( \overline{W}_{\mu\nu}^{1\times 1} )^{ - 3/4 }
        =( 1 - 0.8412 / \beta        )^{ - 3/4 }
        =1.398
\ ,
\end{equation}
where $\overline{W}_{\mu\nu}^{1\times 1}$ is
the value in one-loop perturbation theory~\cite{RG-gauge:IW}.
Quark propagators are solved with the Dirichlet boundary condition
imposed in the time direction for gauge configurations fixed to
the Coulomb gauge.
The wall source defined by (\ref{wall_def.eq})
is set at $t_0=12$ which is sufficient to
avoid effects from the temporal boundary.

The lattice cutoff was estimated as
$1/a=1.207(12)\ {\rm GeV}$ ($a = 0.1632(16)\ {\rm fm}$) from
$m_\rho$~\cite{CP-PACS:LHM}.
The lattice sizes (the numbers of configurations in parentheses) are
$16^3\times 80$ $(1200)$,
$20^3\times 80$ $(1000)$ and
$24^3\times 80$ $(506 )$,
which correspond to the lattice extent
$2.61$,
$3.26$ and
$3.92\ {\rm fm}$ in physical units.
Five quark masses are chosen to give
$m_\pi^2 = $
$0.273$,
$0.351$,
$0.444$,
$0.588$ and
$0.736 \ {\rm GeV}^2$.
The numbers of positions that give independent wave functions are
$165$,
$286$ and
$455$ for $L=16$, $20$ and $24$.
%
%
\section{ Results }
\label{Sec:Results}
%
%
\subsection{ Wave functions }
\label{Sec:Results_zan24}
\begin{figure}[h]
\begin{center}
\includegraphics[width=160mm]{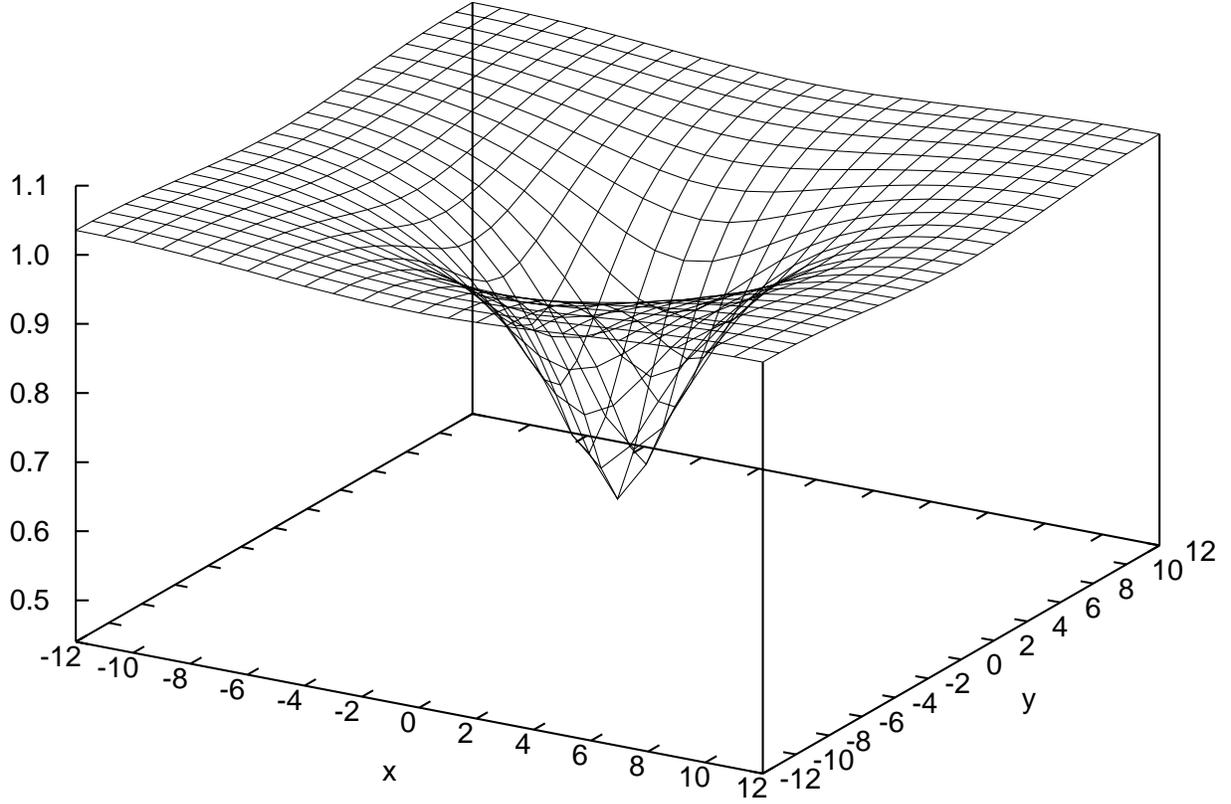}
\caption{
Two-pion wave function $\phi(\vec{x};k)$ on $24^3$ lattice
on $(t,z)=(52,0)$ plane for $m_\pi^2=0.273\ {\rm GeV}^2$.
The reference vector is set at
$\vec{x}_0=(7,5,2)$ ($x_0=|\vec{x}_0|=8.832$).
}
\label{FIG-RXN.2480.K4.52.0088_gP.fig}
\end{center}
\newpage
\end{figure}

The two-pion wave function calculated on the $24^3$ lattice
is exemplified in Fig.~\ref{FIG-RXN.2480.K4.52.0088_gP.fig}
on the $(t,z)=(52,0)$ plane for $m_\pi^2=0.273\ {\rm GeV}^2$,
with the reference position in (\ref{method_wfunc.eq})
fixed at $\vec{x}_0=(7,5,2)$ ($x_0=|\vec{x}_0| = 8.832$).
The statistical errors are negligible in the scale of the figure.

\begin{figure}[h]
\begin{center}
\includegraphics[width=140mm]{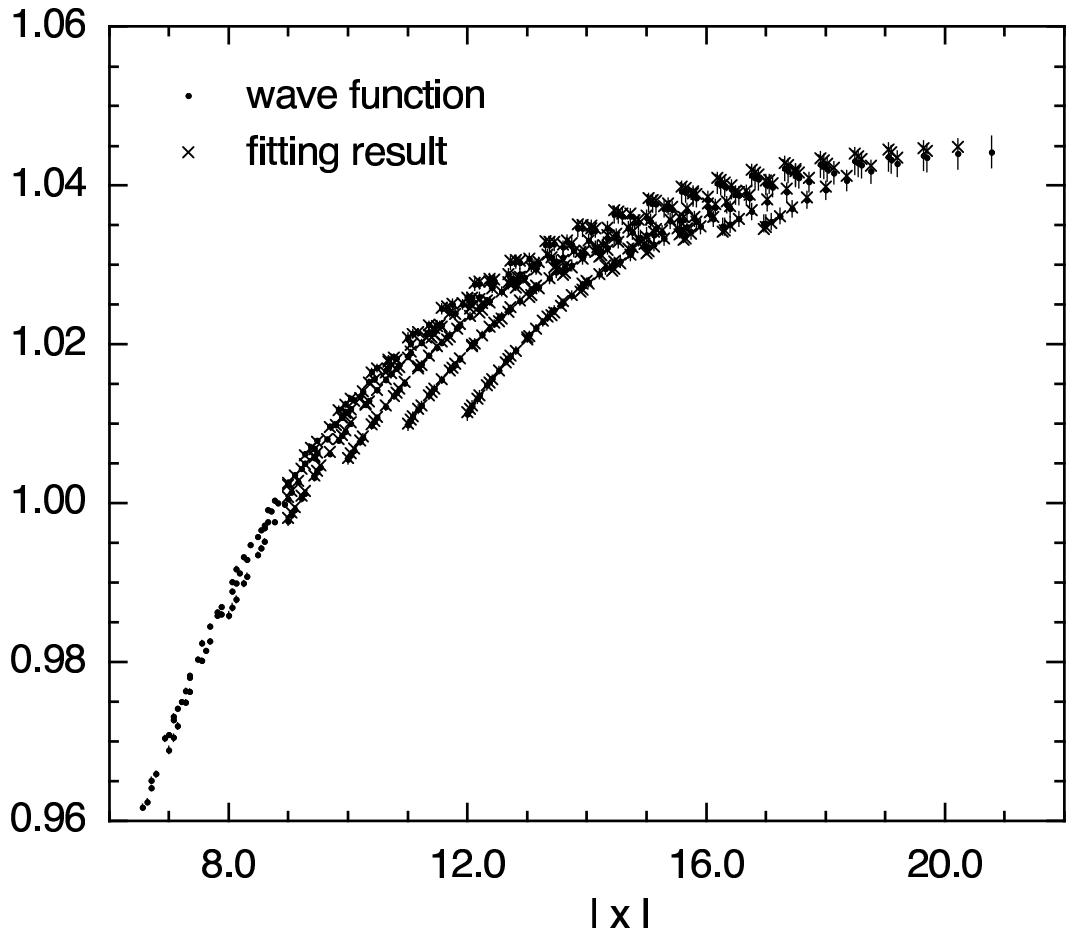}
\caption{
Two-pion wave function $\phi(\vec{x};k)$ on $24^3$ lattice
at $t=52$ for $m_\pi^2=0.273\ {\rm GeV}^2$.
Horizontal axis is $x=|\vec{x}|$.
Filled symbols are data points
and cross symbols are results of fitting
with $G(\vec{x};k)$.
}
\label{ZAN-RXNSQ.2480.K4.52.1.0090.0204.0089.fig}
\end{center}
\newpage
\end{figure}

The same wave function is shown in
Fig.~\ref{ZAN-RXNSQ.2480.K4.52.1.0090.0204.0089.fig}
as a function of $x=|\vec{x}|$ for independent data points.
The branching of the curve seen in the figure
indicates that the wave function does not represent a pure $S$-wave.
This can be understood by the consideration in what follows.
Let $f(x)$ be a function depending only on $x=|\vec{x}|$
for $\vec{x}=[-L/2,L/2]^3$.
The first derivative is given by
\begin{equation}
  \nabla f(x) = \frac{\vec{x}}{x} \frac{\rm d}{ {\rm d}x } f(x)
\ .
\end{equation}
Thus $f(x)$ satisfies the boundary condition
only when ${\rm d}f(x)/{\rm d}x =0$ at the boundary
where at least one component of the vector $\vec{x}$ takes $\pm L/2$.
This also means ${\rm d}f(x)/{\rm d}x =0$ for $x\ge L/2$
from symmetry under the cubic group.
The wave function for the scattering system
generally does not satisfy this.
Hence it cannot be a pure $S$-wave function,
but receives contributions from the states with angular momenta $l>0$.
We expect that
the wave function that belongs to the ${\bf A_1^+}$ representation
contains $j_l(kx)$ with $l\ge 4$,
but not $n_l(kx)$ with $l\ge 4$,
because $\delta_l(k)$ is small for $l\ge 4$ for the two-pion ground state.
This is supported by our results as shown latter.

\begin{figure}[h]
\begin{center}
\includegraphics[width=160mm]{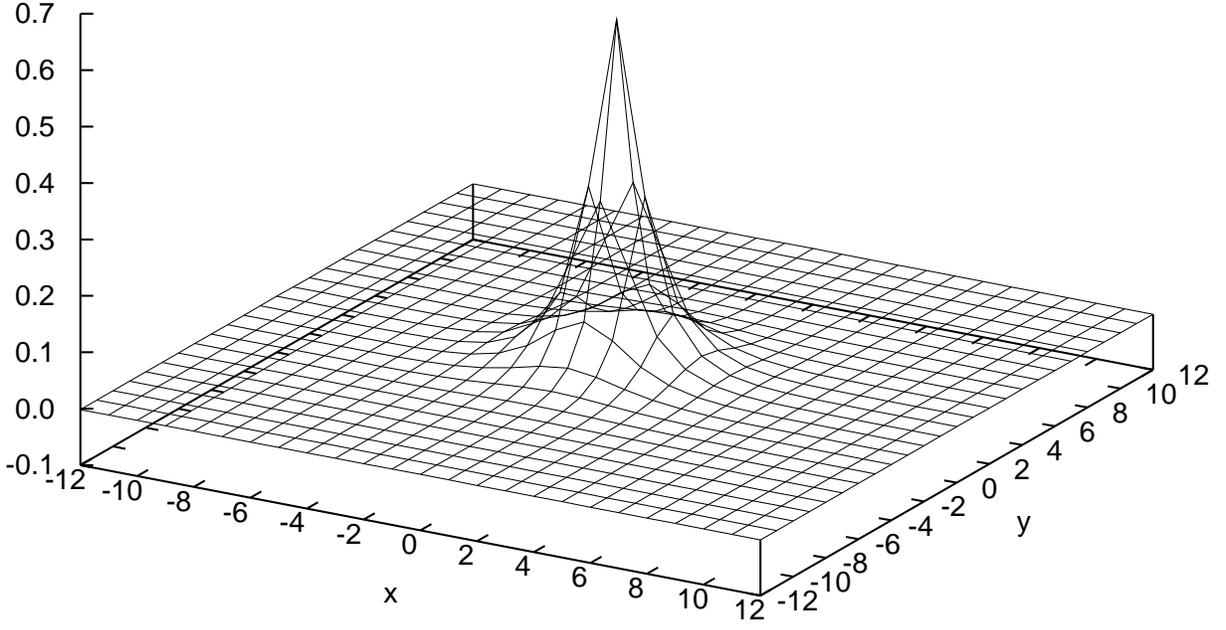}
\caption{
$V(\vec{x};k)$ in unit of $1/a^2$ on $24^3$ lattice
on $(t,z)=(52,0)$ plane for $m_\pi^2=0.273\ {\rm GeV}^2$.
}
\label{POT-WXSP.2480.K4.52_gP.fig}
\end{center}
\newpage
\end{figure}

We now consider the two-pion interaction from the ratio :
\begin{equation}
  V(\vec{x};k)
     = \frac{ \triangle \phi(\vec{x};k) }{ \phi(\vec{x};k) }
\ .
\label{def_ratio_V.eq}
\end{equation}
Here we adopt the naive numerical Laplacian on the lattice,
\begin{equation}
  \triangle f(\vec{x})
   =  \sum_\mu \bigl[
            f( \vec{x} + \hat{\mu} )
        -   f( \vec{x} - \hat{\mu} )
        + 2 f( \vec{x} )
   \bigr]
\ ,
\end{equation}
since $k^2$ is very small
and the choice of the numerical Laplacian is not important for a large $x$.
Away from the two-pion interaction range,
{\it i.e.} $x>R$, we expect that
$V(\vec{x};k)$ is independent of $\vec{x}$ and equals to $-k^2$.
In Fig.~\ref{POT-WXSP.2480.K4.52_gP.fig}
$V(\vec{x};k)$ is plotted for the same parameters as for
Fig.~\ref{FIG-RXN.2480.K4.52.0088_gP.fig}.
The statistical errors are again negligible.
We find a very clear signal and $V(\vec{x};k)$ seems to be constant for $x>8$.
We observe a strong repulsive interaction at the origin
consistent with the negative
scattering length of the $I=2$ two-pion system.

\begin{figure}[h]
\begin{center}
\includegraphics[width=160mm]{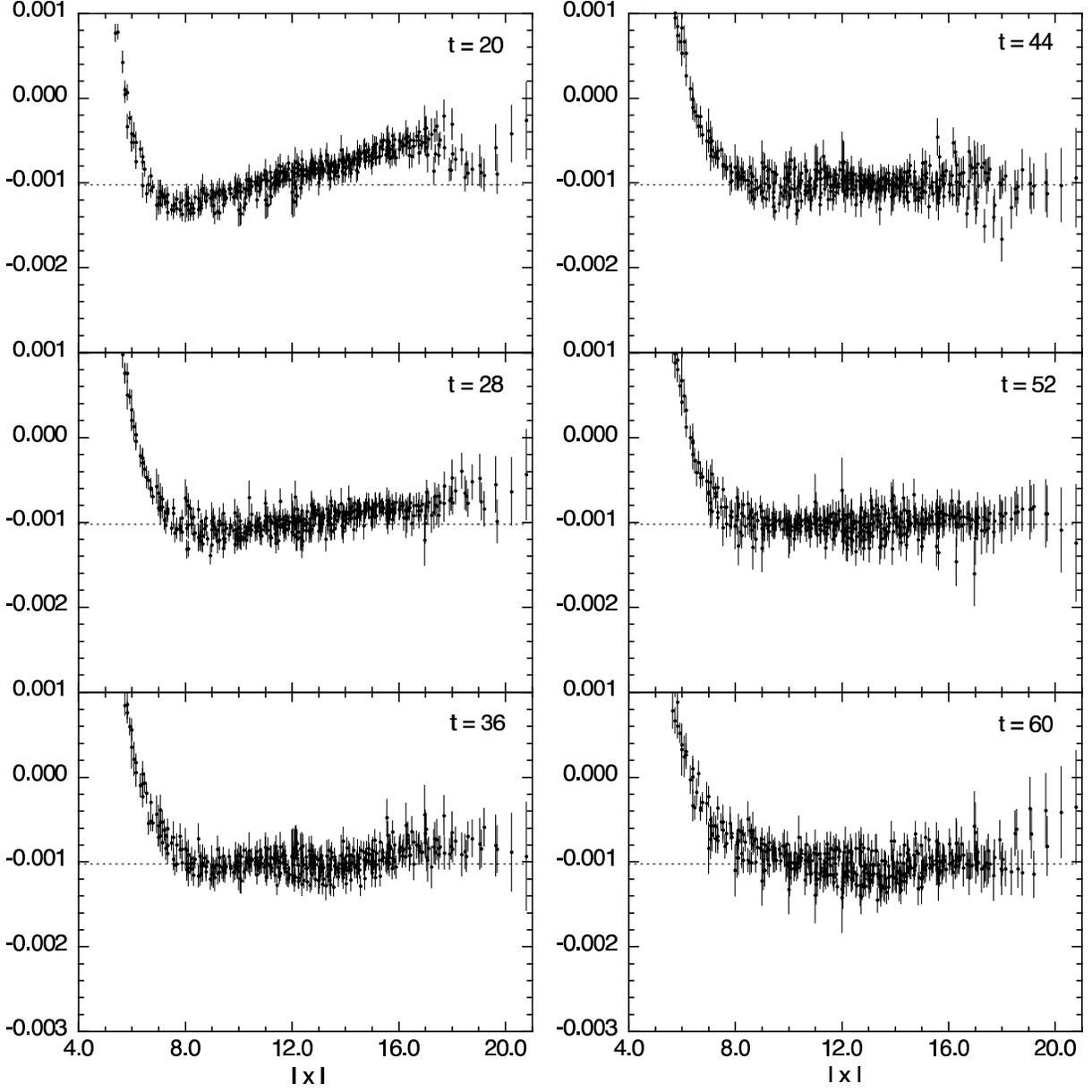}
\caption{
Time dependence of $V(\vec{x};k)$ in unit of $1/a^2$
on $24^3$ lattice for $m_\pi^2=0.273\ {\rm GeV}^2$.
Horizontal axis is $x=|\vec{x}|$.
We plot a line at $-k^2$
estimated from the two-pion time correlator.
}
\label{POT-WXSP.2480.K4.XX.fig}
\end{center}
\newpage
\end{figure}

The time dependence is shown
in Fig.~\ref{POT-WXSP.2480.K4.XX.fig} for the same parameters,
where the abscissa is $x=|\vec{x}|$
and only independent data are plotted.
We draw a line of $-k^2$
estimated from $\Delta W$ of (\ref{R_t_dep.eq})
using the normalized two-pion time correlator $R(t)$.
We see that $V(\vec{x};k)$ approaches a value consistent with
$-k^2$ for a large $x$.
$V(\vec{x};k)$ depends only on $x$
and does not depend on $t$ in so far as $t\ge 44$.
We also confirm that the wave function $\phi(\vec{x};k)$
does not vary when $t$ is varied for $t\ge 44$.

We now consider the two-pion interaction range $R$.
In quantum field theory
the wave function does not strictly satisfy
the Helmholtz equation (\ref{AsumeU.eq})
even for the large $x$ region in a large volume lattice.
Hence, with $k^2$ obtained from the asymptotic time behavior
of the two-pion time correlator,
$V(\vec{x};k)+k^2$ shows a small tail at a large $x$.

We may take the wave function as satisfying the Helmholtz equation,
if $V(\vec{x};k)+k^2$ is sufficiently small compared with $k^2$.
In the present work we take an operational definition for the
range $R$ as the scale,
where
\begin{equation}
  U(\vec{x};k)
       = \frac{ V(\vec{x};k) + k^2 }{ k^2 }
       = \frac{ ( \triangle + k^2 ) \phi(\vec{x};k) }{ k^2 \phi(\vec{x};k) }
\label{def_ratio_U.eq}
\end{equation}
gets small enough so that it is buried
into the statistical error,
with the expectation that
the systematic error for the scattering length
from the interaction of tails of the wave function becomes smaller
than statistical errors in the resulting scattering length
with this definition.~\footnote{
The radius $R$ thus defined has no direct relevance
to the physical scale such as 
an effective range $r_0$
defined by $k/\tan\delta_0(k)=1/a_0 + r_0 k^2/2 + O(k^4)$.
It becomes larger as the statistical accuracy 
of $k^2$ and $\phi(\vec{x};k)$ increases. 
We needed this somewhat artificial criterion to define $R$, 
unless otherwise we must appeal to some effective models.
We also note that the rigorous estimation of the effects of
the tail on the scattering length is not possible, unless
$U_k(\vec{x},\vec{y})$ for all $\vec{x}$ and 
$\vec{y}$ or the two-pion scattering amplitude off the mass shell 
for all energies is known. 
}

\begin{figure}[h]
\begin{center}
\includegraphics[width=160mm]{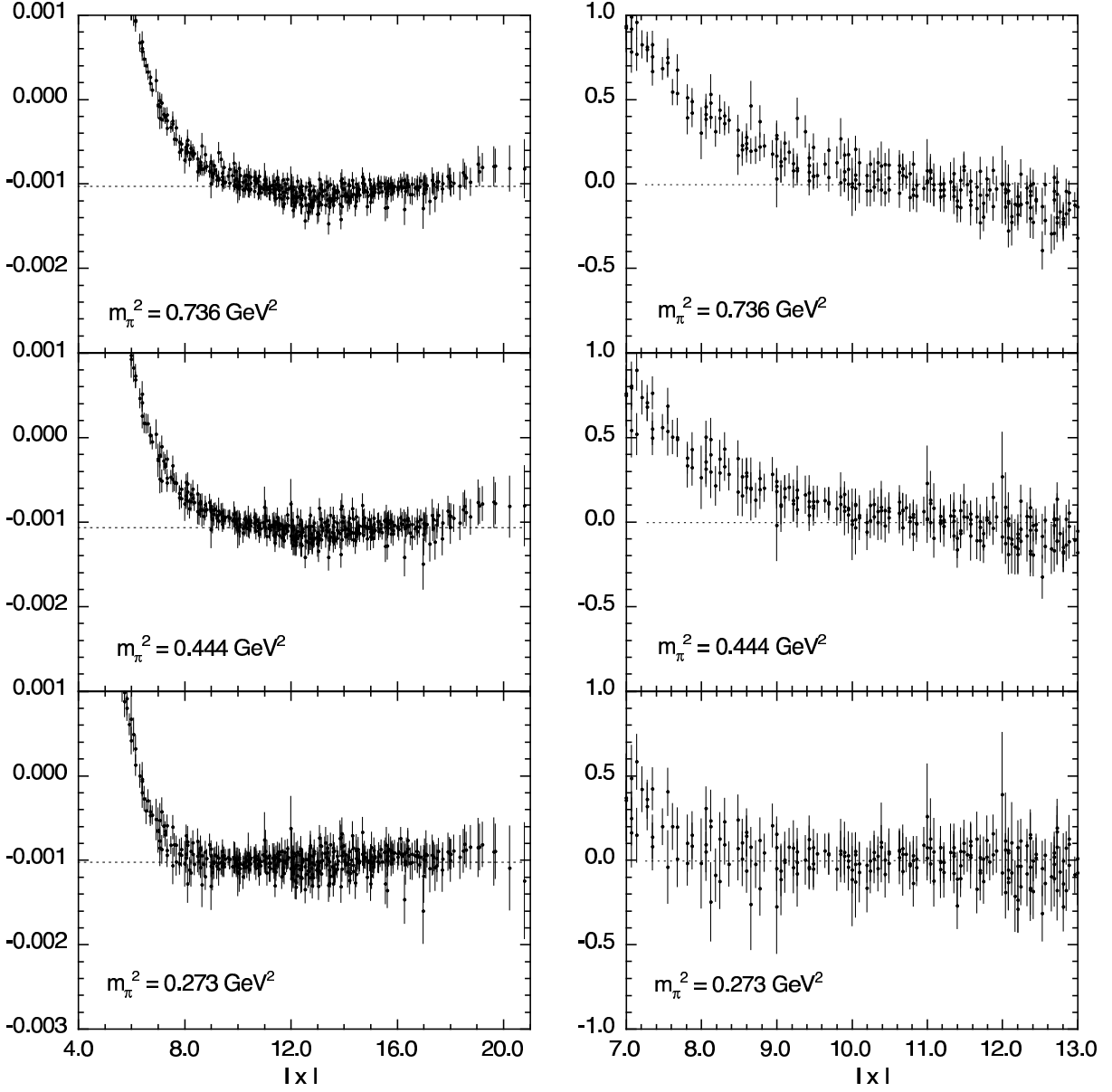}
\caption{
$V(\vec{x};k)$ in unit of $1/a^2$ (left side) and $U(\vec{x};k)$ (right side)
on $24^3$ lattice at $t=52$ for several quark masses.
Horizontal axis is $x=|\vec{x}|$.
We plot a line at $-k^2$
obtained from the two-pion time correlator in left side of figures.
}
\label{POT-WX.2480.KX.52X.fig}
\end{center}
\newpage
\end{figure}

The function $U(\vec{x};k)$
is displayed in Fig.~\ref{POT-WX.2480.KX.52X.fig} at $t=52$,
together with $V(\vec{x};k)$ defined by (\ref{def_ratio_V.eq}),
for which the line of $-k^2$
estimated from the two-pion time correlator is also drawn.
These figures show that
$V(\vec{x};k)\sim -k^2$ and $U(\vec{x};k)\sim 0$ for $x>R$
within the statistical error for all quark masses.
We find $R \sim 10$ ($1.6\ {\rm fm}$) for
the heaviest quark mass, $m_\pi^2=0.736\ {\rm GeV}^2$.
This stands for the largest $R$ we obtained.
This result signifies that
the necessary condition for L\"uscher's formula (\ref{AsumeU.eq})
is satisfied on the $24^3$ lattice
for all our quark masses $m_\pi^2=0.273-0.736\ {\rm GeV}^2$
with the current statistics of simulations.
%
%
\subsection{ Scattering lengths }
\label{Sec:Results_SCL}
We may estimate the scattering length from the wave function
with two alternative methods :
\renewcommand{\labelenumi}{\arabic{enumi}.}
\begin{enumerate}
\item \label{method_I}
We extract $k^2$ by fitting the asymptotic value of $V(\vec{x};k)$
to a constant
and obtain the scattering length
by substituting $k^2$ into~(\ref{Lusher.eq}).
The resulting $k^2$ and the scattering length are given
in Table~\ref{SCL-POT-ZAN-RXPA.2480.KX.table}
in the column labeled with ``from $V$''.
We choose $t=52$ and the fitting range $x_m \le x \le \sqrt{3}L/2$
(maximum value of $x$ for $L^3$ lattice).
The energy shift $\Delta W$ is calculated from $k^2$
using (\ref{eq:deltaW}).
\item \label{method_II}
$k^2$ is obtained
by fitting the wave function $\phi(\vec{x};k)$
with periodic Green function $G(\vec{x};k)$
defined by (\ref{Luscher_wave_Glm.eq}),
taking $k^2$ and an overall constant as free parameters.
An example of fitting is illustrated in
Fig.~\ref{ZAN-RXNSQ.2480.K4.52.1.0090.0204.0089.fig}
at $t=52$ for $m_\pi^2=0.273\ {\rm GeV}^2$,
where the values from fits are shown with cross symbols.
The fitting range is the same as that for method~\ref{method_I}.
The method for numerical evaluation of $G(\vec{x};k)$
is discussed in Appendix~\ref{appe:zeta}.
The fit works well, meaning that
the contributions of $G_{lm}(\vec{x};k)$
and $\delta_l(k)$ with $l\ge 4$ are negligible as expected.
This point is confirmed by fitting with
the function including $G_{40}(\vec{x};k)$.
The results are given
in Table~\ref{SCL-POT-ZAN-RXPA.2480.KX.table} (labeled ``from $\phi$'').
\end{enumerate}

\begin{figure}[h]
\begin{center}
\includegraphics[width=140mm]{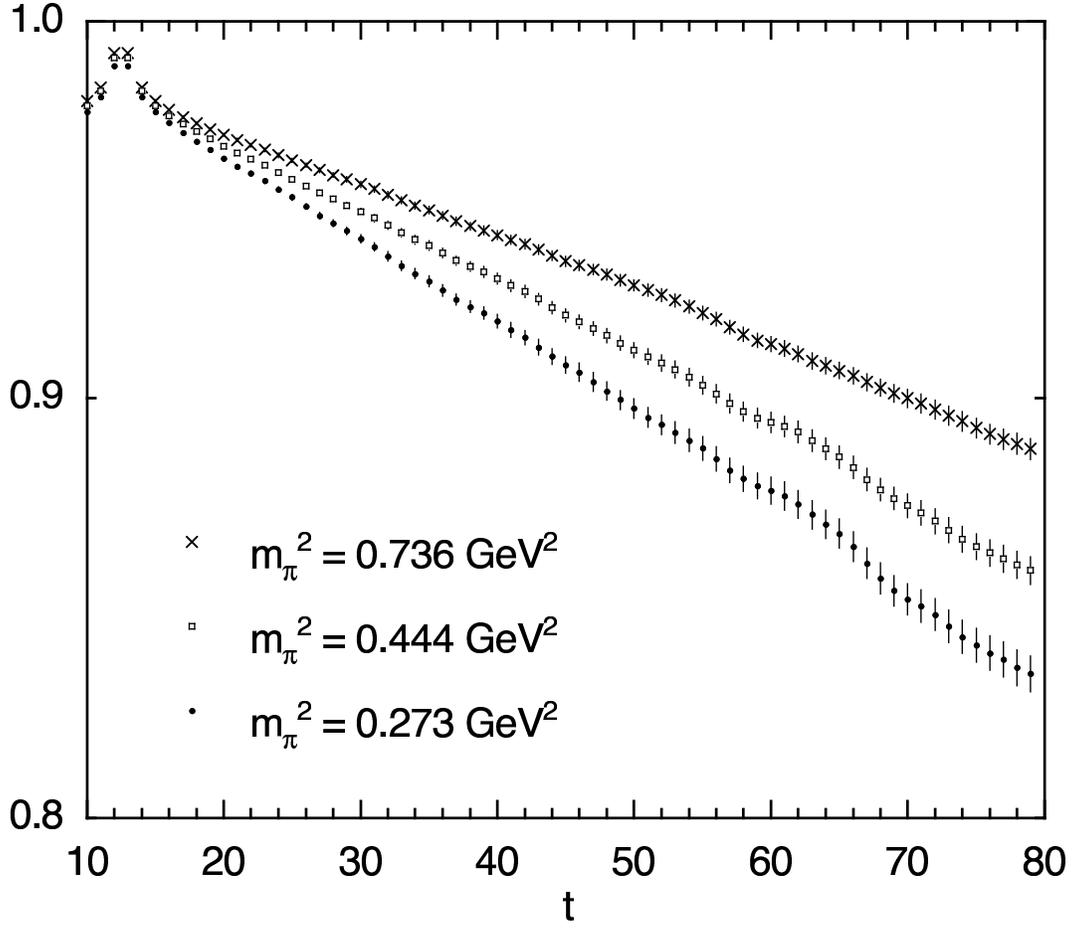}
\caption{
Normalized two-pion correlator $R(t)$ on $24^3$ lattice
for $m_\pi^2 = 0.273$ (lightest), $0.44$ and $0.736\ {\rm GeV}^2$ (heaviest).
Scale of vertical axis is log-scale.
}
\label{SCL-RXPA.2480.KX.fig}
\end{center}
\newpage
\end{figure}

\begin{figure}[h]
\begin{center}
\includegraphics[width=100mm]{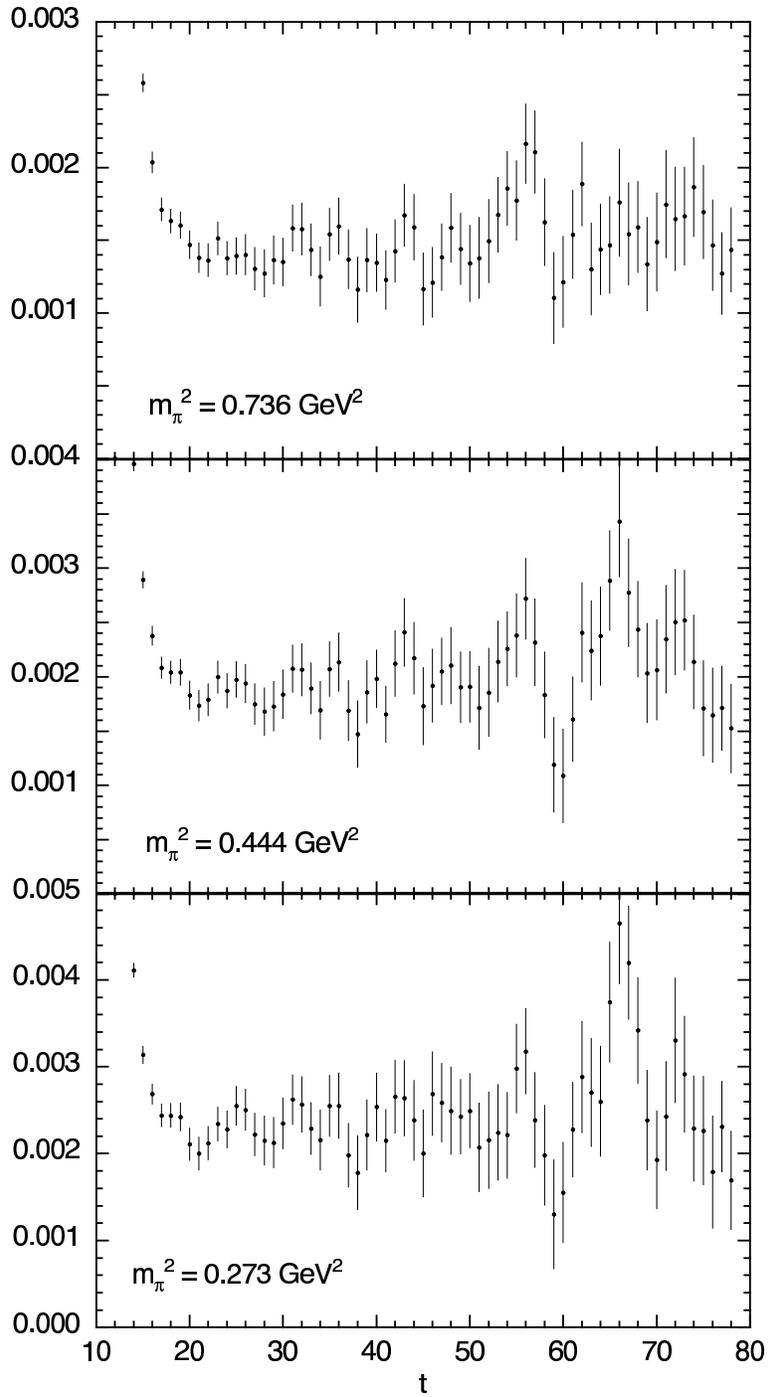}
\caption{
Effective mass for $R(t)$ on $24^3$ lattice in unit of $1/a$
at $m_\pi^2=0.273$ (lightest), $0.444$ and $0.736\ {\rm GeV}^2$ (heaviest).
}
\label{SCL-RXPAL.2480.KX.fig}
\end{center}
\newpage
\end{figure}

\begin{figure}[h]
\begin{center}
\includegraphics[width=140mm]{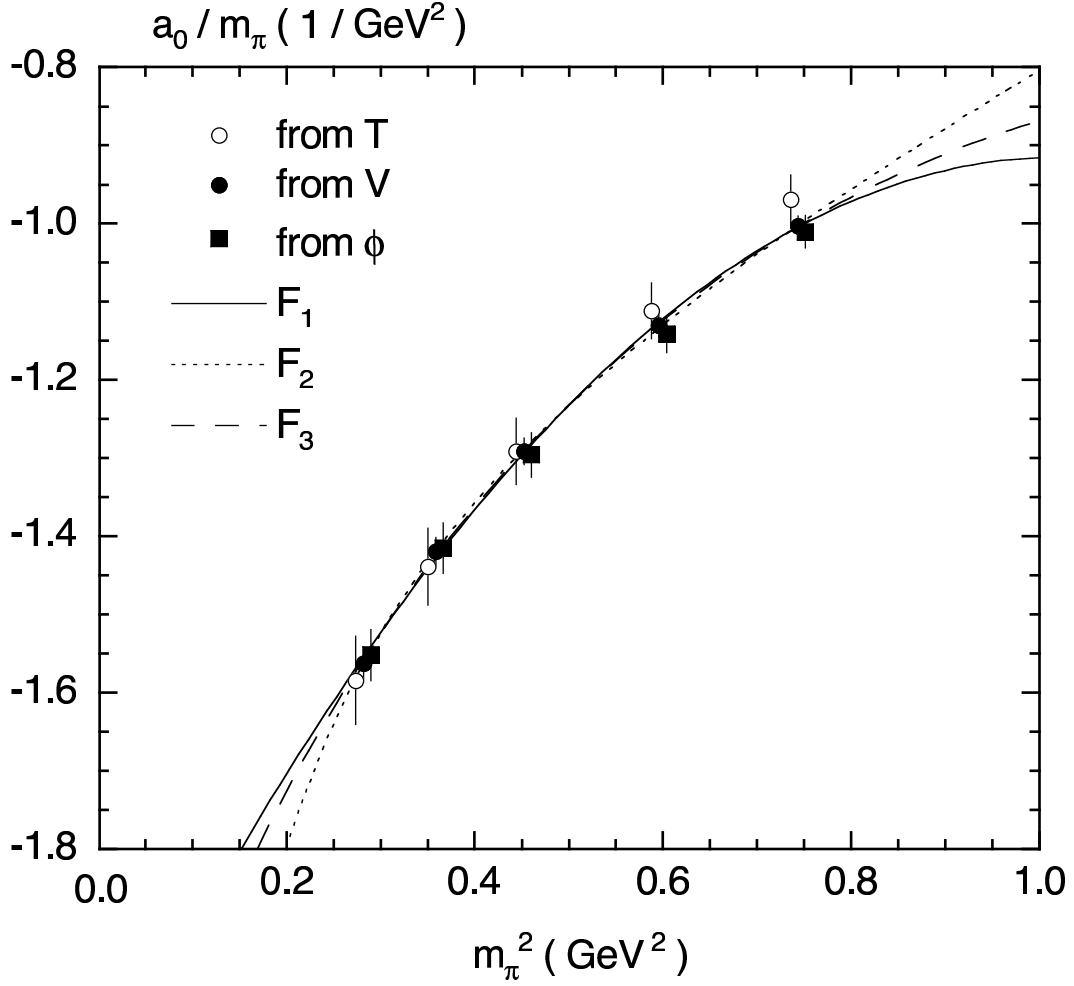}
\caption{
Pion scattering length $a_0/m_\pi\ (1/{\rm GeV}^2)$ on $24^3$ lattice
obtained from the two-pion time correlator (``from $T$''),
the constant fit of $V(\vec{x};k)$ (``from $V$'')
and the fitting the wave function (``from $\phi$'').
Results of the chiral extrapolation
of $a_0/m_\pi$ obtained from $V(\vec{x};k)$ (``from $V$'')
with the three fitting functions are also plotted lines.
}
\label{FR_2480.fig}
\end{center}
\newpage
\end{figure}

We compare the scattering lengths obtained from the wave function
with those from the conventional method of
using the two-pion time correlator.
We plot the normalized two-pion correlator $R(t)$ of (\ref{R_t_def.eq})
in Fig.~\ref{SCL-RXPA.2480.KX.fig},
which shows a clear signal that decreases exponentially in $t$.
This means the absence of the singular-value solution for this volume,
or otherwise $R(t)$ should approach some constant.
The effective masses for $R(t)$
presented in Fig.~\ref{SCL-RXPAL.2480.KX.fig} show
the plateau over $t \sim 22-70$.
Rows indicated with the label ``from T''
in Table~\ref{SCL-POT-ZAN-RXPA.2480.KX.table}
present the values of $\Delta W$ obtained
by a single exponential fitting for $t=24-68$,
$k^2$ estimated from $\Delta W$,
and the scattering length from L\"uscher's formula (\ref{Lusher.eq}),
using $k^2$ thus estimated and $g_{00}(k)$
calculated by the numerical method discussed in~\cite{CP-PACS:phsh}.
We compare the scattering length
obtained from three methods in Fig.~\ref{FR_2480.fig}.
The three methods give consistent results within statistical errors.
We observe that the statistical errors for those
from our new methods (``from $V$'' and ``from $\phi$'')
are significantly smaller
than those from the two-pion time correlator (``from $T$'').
This feature was experienced
in the two-dimensional statistical models~\cite{Wisz:stat}.
Our analysis so far is made at $t=52$, but we checked that the
results are independent of the choice of $t$
in so far as $t\ge 48$.

When one wants to obtain scattering lengths for the physical pion mass,
there is yet an important problem of the chiral extrapolation.
In Fig.~\ref{FR_2480.fig}
we carry out the chiral extrapolation
of $a_0/m_\pi$ using the data from method~{\ref{method_I}}
({\it i.e.} ``from $V$'')
by assuming three fitting forms :
\begin{eqnarray}
    F_1( m_\pi^2 ) & =
           & A_1 + B_1 \cdot m_\pi^2 + C_1 \cdot m_\pi^4   \cr
    F_2( m_\pi^2 ) & =
           & A_2 + B_2 /     m_\pi^2 + C_2 \cdot m_\pi^2   \cr
    F_3( m_\pi^2 ) & =
           & A_3 \biggl( 1 + B_3 \cdot m_\pi^2 \cdot \log ( m_\pi^2 / C_3 ) \biggr)
\ \ .
\end{eqnarray}
The form of $F_2(m_\pi^2)$ is motivated
from chiral symmetry breaking of the Wilson fermion
and quenching effects, as suggested from quenched chiral perturbation
theory~\cite{Bernard-Golterman:QCHPT,Colangelo-Pallante:QCHPT}.
$F_3(m_\pi^2)$ is the form predicted by chiral perturbation theory (CHPT)
for full QCD in the one loop order.
$F_1(m_\pi^2)$ is a simple polynomial.
We see that the three functions fit the data equally well
and we cannot distinguish among them.
The chiral limit of $a_0/m_\pi(=A_j)$, however,
depends sizably on the choice of the fitting forms :
\begin{eqnarray}
    a_0 / m_\pi\ (1/{\rm GeV}^2)
       & = \ -2.117(83)  & \mbox{ from $F_1(m_\pi^2)$ } \cr
       & = \ -1.29(15)\  & \mbox{ from $F_2(m_\pi^2)$ } \cr
       & = \ -2.39(16)\  & \mbox{ from $F_3(m_\pi^2)$ }
\ \ .
\end{eqnarray}
We cannot reduce this large systematic errors arising
from the choice of fitting forms,
unless simulations are made close to the physical pion mass or the fitting
form is theoretically constrained. We add that the prediction from CHPT
is $a_0/m_\pi = -2.265(51)\ (1 / {\rm GeV}^2)$~\cite{CHPT_a0_cont:CHPT}.
%
%
\subsection{ Results on small lattices }
\label{Sec:Results_zan_small}
We carry out the same analysis for $20^3$ and $16^3$ lattices
to study the dependence on the finite lattice size,
with the numerical results presented in
Tables~\ref{SCL-POT-ZAN-RXPA.2080.KX.table} and
       \ref{SCL-POT-ZAN-RXPA.1680.KX.table}.

Our analysis on the  $24^3$ lattice shows that
the two-pion interaction range is at most $R\sim 10$ ($1.6\ {\rm fm}$)
which happens for the heaviest quark mass, $m_\pi^2 = 0.736\ {\rm GeV}^2$,
so that L\"uscher's formula (\ref{Lusher.eq}) is safely applied
for the $24^3$ lattice for all our quark masses
with the present statistical accuracy.
This appears to indicate that $20^3$ is needed and $16^3$ may be too small.
This is not necessarily true, however,
since $R$ depends on the quark mass and the momentum $k^2$
which strongly depends on the lattice volume,
as is seen by comparing the relevant entries in the three tables.

\begin{figure}[h]
\begin{center}
\includegraphics[width=160mm]{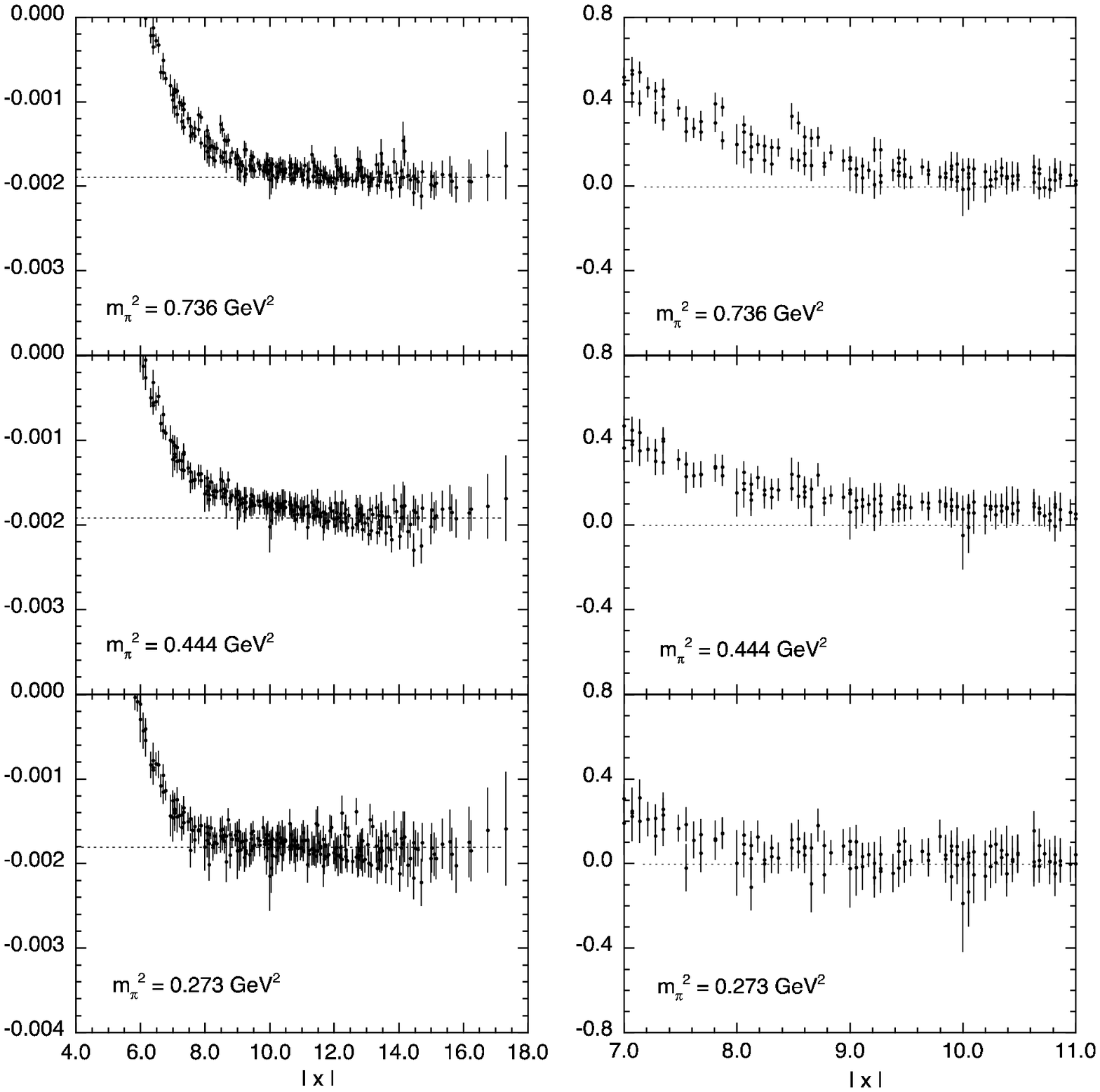}
\caption{
$V(\vec{x};k)$ in unit of $1/a^2$ (left side)
and $U(\vec{x};k)$ (right side)
on $20^3$ lattice
at $t=52$ for several quark masses.
In left side of figures we also plot a line at $-k^2$
obtained from the two-pion time correlator.
Horizontal axis is $x=|\vec{x}|$.
}
\label{POT-WX.2080.KX.52X.fig}
\end{center}
\newpage
\end{figure}

\begin{figure}[h]
\begin{center}
\includegraphics[width=160mm]{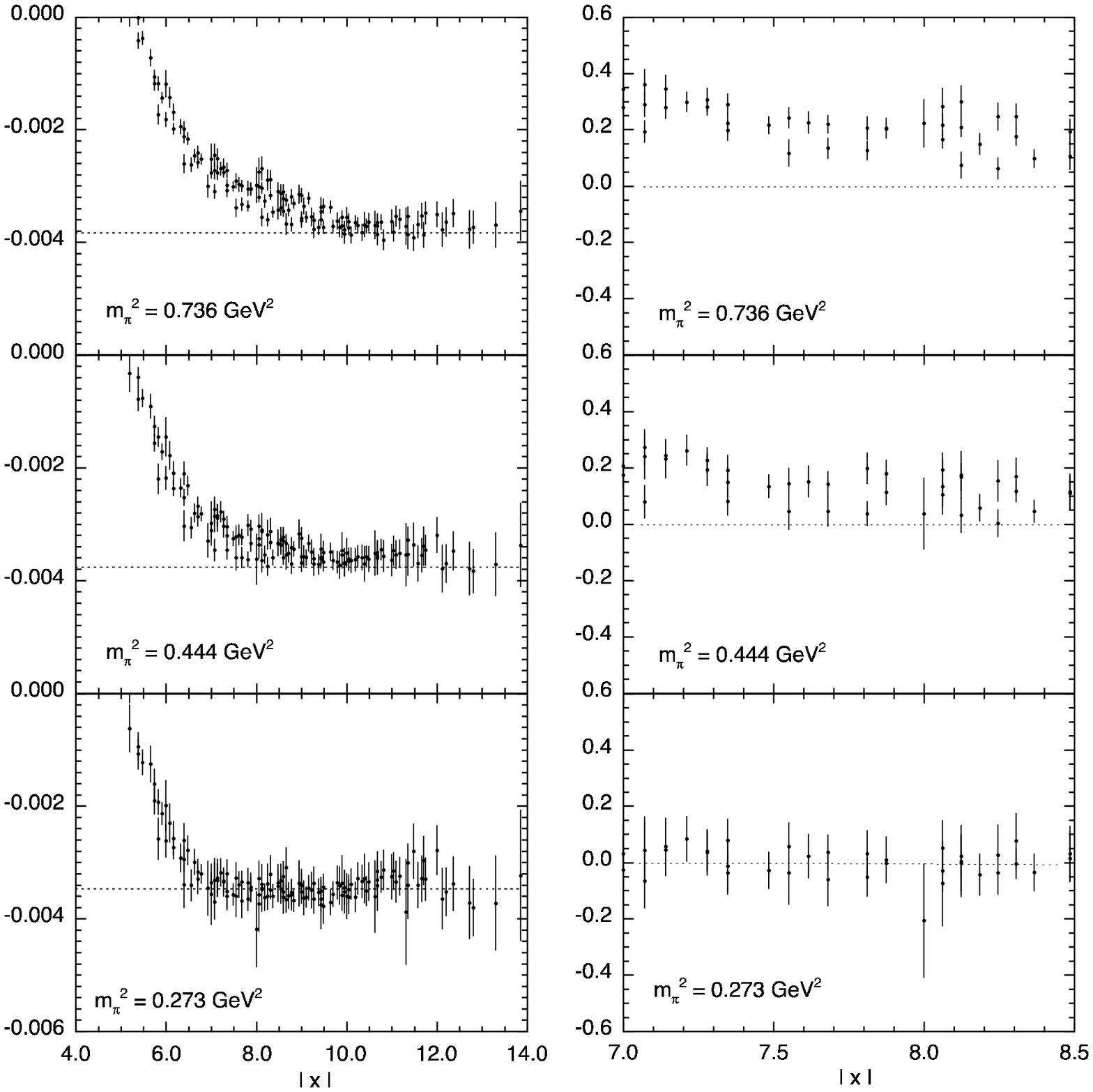}
\caption{
$V(\vec{x};k)$ in unit of $1/a^2$ (left side)
and $U(\vec{x};k)$ (right side)
on $16^3$ lattice
at $t=52$ for several quark masses.
In left side of figures
we also plot a line at $-k^2$
obtained from the two-pion time correlator.
Horizontal axis is $x=|\vec{x}|$.
}
\label{POT-WX.1680.KX.52X.fig}
\end{center}
\newpage
\end{figure}

To investigate the lattice size dependence of the interaction range,
we plot $V(\vec{x};k)$ and $U(\vec{x};k)$
for the $20^3$ lattice in Fig.~\ref{POT-WX.2080.KX.52X.fig}, and
for the $16^3$ lattice in Fig.~\ref{POT-WX.1680.KX.52X.fig}, both at $t=52$.
We cannot clearly observe a region $x<L/2$ where $U(\vec{x};k)\sim 0$
for heavier quarks,
while such a region within statistics is visible for lighter quarks.
The scattering lengths are calculated for the latter cases,
{\it i.e.} at the two lighter quark masses
$m_\pi^2=0.273$ and $0.351\ {\rm GeV}^2$ on the $20^3$ lattice,
and only at the lightest quark mass
$m_\pi^2 = 0.273\ {\rm GeV}^2$
on the $16^3$ lattice.

\begin{figure}[h]
\begin{center}
\includegraphics[width=140mm]{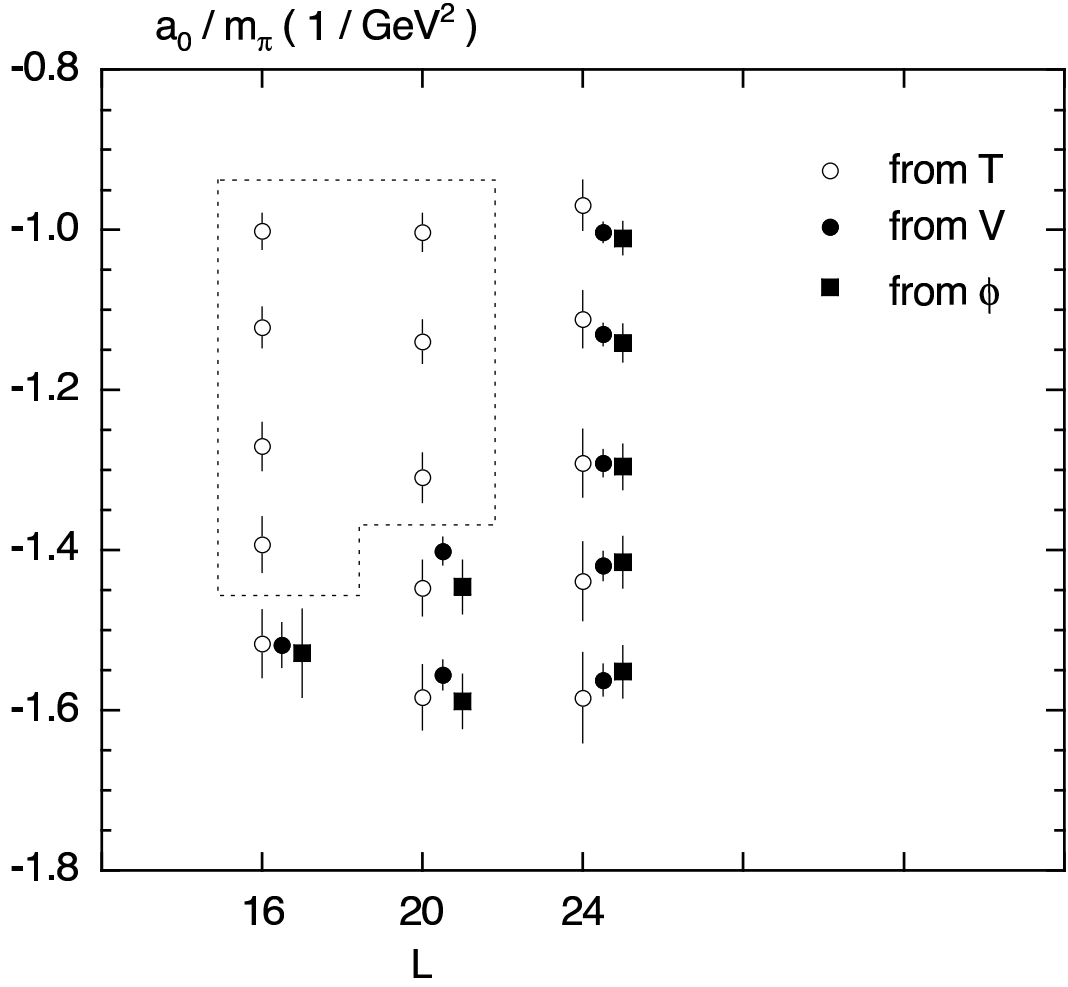}
\caption{
Volume dependence of pion scattering length $a_0/m_\pi \ (1/{\rm GeV}^2)$
for five quark masses
obtained from the time correlator (``from $T$'')
and the wave function (``from $V$'' and ``from $\phi$'').
}
\label{comp.fig}
\end{center}
\newpage
\end{figure}

Our compilation of the the scattering lengths,
{\it i.e.} those obtained on the three lattice volumes
for five quark masses with three different methods,
is depicted in Fig.~\ref{comp.fig}.
Data points encircled by dotted lines are
those from the two-pion time correlator
for the case for which we cannot clearly find
a region $x<L/2$ where the two-pion interaction vanishes.
We do not find a significant volume dependence,
however, for all quark masses
including the case
where the necessary condition for L\"uscher's formula is not satisfied.
The effects of deformation of the two-pion interaction
due to finite volume effects on $\Delta W$ of the two-pion system
are apparently small compared with the statistical error.
We emphasize, however, that the reliability of these data
is guaranteed only after
we obtain the results on the $24^3$ lattice
where the necessary condition for L\"uscher's formula is satisfied.
%
%
\section{ Conclusions }
\label{Sec:Conclusions}
We have shown in this work
that calculations of the two-pion wave function
for the ground state of the $I=2$ $S$-wave two-pion system is feasible.
We have investigated the validity of
the necessary condition for L\"uscher's formula
and have found that it is satisfied for $L\ge 24$ ($3.92\ {\rm fm}$)
for $m_\pi^2 = 0.273-0.736\ {\rm GeV}^2$.
We have demonstrated that
the scattering length can be extracted from the wave function
with smaller statistical errors
than from the two-pion time correlator
which has been used in the studies to date.

We have observed no significant volume dependence
for the scattering lengths
obtained from the two-pion time correlator at least
for $L\ge 16$ ($2.61\ {\rm fm}$),
in spite of the fact that the necessary condition for the formula
is not satisfied in some cases for $L=16$ and $20$.
The effects of deformation of the two-pion interaction
due to finite size effects on the energy eigenvalues of the two-pion
are likely small compared with the statistical error.

The present work opens the possibility to reduce statistical errors
in the scattering phase shift with modest statistics.
In this case, however, we have to concern
with the contamination from inelastic scattering.
This effect is probably negligible for the ground state
of the two-pion as in this work,
but it may be important for the momentum excitation states,
whose energies are close to the inelastic threshold.
One may investigate effects of inelastic scattering
by evaluating $I(\vec{x})$ in (\ref{BS_def_infinit2.eq})
in Appendix~\ref{appe:Luscher-RFT}
from calculations of $\phi(\vec{x};k)$
and $\langle \pi(\vec{p}) | \pi(-\vec{x}/2) | \pi\pi; k \rangle$.
This would also clarify the error associated with our neglect of the
inelastic channel, but the work is deferred to the future.

Another implication of the present work
is the feasibility to calculate the decay width of the $\rho$ meson
through studies of the $I=1$ two-pion system.
While the evaluation of the disconnected diagrams
with a good precision has been a computational problem,
our new method,
investigating the scattering system
from the wave function of multi-hadron states
in which the energy eigenvalue is extracted
from the wave function at a single time slice,
could lend a tactics that
can be used to evaluate such complicated diagrams
with a modest cost.
%
%
\section*{Acknowledgments}
This work is supported in part by Grants-in-Aid of the Ministry of Education
(Nos.
12304011, 
12640253, 
13135204, 
13640259, 
13640260, 
14046202, 
14740173, 
15204015, 
15540251, 
15540279, 
15740134  
).
The numerical calculations have been carried out
on the parallel computer CP-PACS.
%
%
\newpage
\appendix
%
%
\section{ L\"uscher's formula from Bethe-Salpeter wave function }
\label{appe:Luscher-RFT}
We discuss the derivation of L\"uscher's formula (\ref{Lusher.eq})
from Bethe-Salpeter (BS) wave function
in quantum field theory~\cite{LMST:kpp}.
All considerations are made in Minkowski space,
but these are not changed even in Euclidian space.

We consider the BS function in the infinite volume defined by
\begin{equation}
  \phi_{\infty} (\vec{x};\vec{k}) =
      \langle 0 | \pi_1(    \vec{x}/2 )
                  \pi_2(  - \vec{x}/2 )
          | \pi_1(\vec{k}), \pi_2(-\vec{k}) ; {\rm in} \rangle
\ ,
\label{BS_def_infinit.eq}
\end{equation}
where we consider the particle as distinguishable
and denote two distinguishable pions by $\pi_1$ and $\pi_2$.
The state $|\pi_1(\vec{k}),\pi_2(-\vec{k}); {\rm in} \rangle$
is an asymptotic two-pion state
with the momentum $\vec{k}$ and $-\vec{k}$,
and $\pi_1(\vec{x})$ and $\pi_2(\vec{x})$
are the interpolating operators for the pion $\pi_1$ and $\pi_2$
at position $\vec{x}$.
Inserting complete intermediate energy eigenstates
between the two fields,
we obtain
\begin{equation}
  \phi_{\infty} (\vec{x};\vec{k})
 =   \int \frac{ {\rm d}^3p }{ (2\pi)^3 2 E_p }
         \sqrt{Z} {\rm e}^{ i \vec{p} \cdot \vec{x} / 2 }
         \langle \pi_1 (\vec{p}) | \pi_2 ( - \vec{x}/2 )
             | \pi_1(\vec{k}), \pi_2(-\vec{k}) ; {\rm in} \rangle
      + I( \vec{x} )
\ ,
\label{BS_def_infinit2.eq}
\end{equation}
where $E_p = \sqrt{ m_\pi^2 + p^2 }$.
$I(\vec{x})$ is contribution of in-elastic scattering
from the states with more than one pions,
whose contributions are expected to be small
for the energy $2 E_k << 4 m_\pi$.
This energy condition is satisfied
for our case of the two-pion ground state,
thus we neglect $I(\vec{x})$ in the following.

We decompose the BS function
into the disconnected and the connected part,
and rewrite them by the LSZ reduction formula as
\begin{equation}
      \frac{  \langle \pi_1(\vec{p}) | \pi_2( - \vec{x}/2 )
                  | \pi_1(\vec{k}), \pi_2(-\vec{k}) ; {\rm in} \rangle  }
           { 2E_p \sqrt{Z} {\rm e}^{ i \vec{p} \cdot \vec{x} /2 } }
    =  (2\pi)^3 \delta^3 ( \vec{ p } - \vec{k} )
    + \frac{ H(p;k) }{ p^2 - k^2 - i\epsilon }
\ ,
\label{14.eq}
\end{equation}
where $H(p;k)$ is related to
the off-shell two-pion scattering amplitude $M(p;k)$ by
\begin{equation}
   H(p;k) = \frac{ E_p + E_k }{ 8 E_p E_k } M(p;k)
\ .
\label{Hpk_eq}
\end{equation}
$M(p;k)$ is defined from the pion 4-point Green function by
\begin{eqnarray}
&&
  {\rm e}^{ - i {\bf q}\cdot{\bf x} }
 \frac{ M(p;k) }{ - {\bf q}^2 + m_\pi^2 - i\epsilon }
 =
     \int {\rm d}^4 z
        \ {\rm d}^4 y_1
        \ {\rm d}^4 y_2 \
          {\bf K}(   {\bf p}  , {\bf z}   )
          {\bf K}( - {\bf k}_1, {\bf y}_1 )
          {\bf K}( - {\bf k}_2, {\bf y}_2 )
\cr
&& \qquad\qquad\qquad
   \qquad\qquad\qquad
   \qquad\qquad
      \langle 0 |
           T \bigl[
               \pi_1 ( {\bf z}   )
               \pi_2 ( {\bf x}   )
               \pi_1 ( {\bf y}_1 )
               \pi_2 ( {\bf y}_2 )
           \bigr]
      | 0 \rangle
\ ,
\label{BS_def_infinitM.eq}
\end{eqnarray}
where all momenta and coordinates
denoted by bold face characters refer to four-dimensional vectors and
\begin{equation}
   {\bf K}( {\bf p} , {\bf z} )
      = \frac{ i }{ \sqrt{Z} }{\rm e}^{ i {\bf p} \cdot {\bf z} }
         ( - {\bf p}^2 + m_\pi^2 )
\ .
\end{equation}
In our case
the momenta in (\ref{BS_def_infinitM.eq}) take
\begin{eqnarray}
   &&         {\bf k}_1 = ( E_k ,  \vec{k} )   \ ,  \quad
              {\bf k}_2 = ( E_k , -\vec{k} )
\cr
   &&         {\bf p}   = ( E_p ,  \vec{p} )  \ ,   \quad
              {\bf q}   = {\bf k}_1 + {\bf k}_2 - {\bf p}
                        = ( 2E_k - E_p , - \vec{p} )
\ .
\end{eqnarray}
${\bf q}$ is generally off-shell momentum
and ${\bf q} = ( E_p, - \vec{p} )$ at on-shell ($E_p=E_k$).
The others are on-shell momenta.

We assume that
the scattering phase shift $\delta_l(k)$ with $l \ge 1$ is negligible,
and regard $H(p;k)$ and $M(p;k)$
as functions of $p=|\vec{p}|$ and $k=|\vec{k}|$.
We also assume regularity for all $p$ and $k$.
$M(p;k)$ and $H(p;k)$ are normalized as
\begin{eqnarray}
  M(k;k) &=& \frac{ 16 \pi E_k }{ k } {\rm e}^{ i \delta_0 (k) } \sin\delta_0 (k) \cr
  H(k;k) &=& \frac{ 4  \pi     }{ k } {\rm e}^{ i \delta_0 (k) } \sin\delta_0 (k)
\label{BS_def_infinitM2.eq}
\end{eqnarray}
at on-shell $p=k$.

Substituting (\ref{14.eq}) into (\ref{BS_def_infinit2.eq}),
we obtain
\begin{eqnarray}
  \phi_{\infty} (\vec{x};\vec{k})
  &=&
      {\rm e}^{ i \vec{k}\cdot\vec{x} }
    +  \int \frac{ {\rm d}^3p }{ (2\pi)^3 }
            \frac{ H(p;k) }{ p^2 - k^2 - i \epsilon }
                   {\rm e}^{ i \vec{p} \cdot \vec{x} }
\cr
  &=& {\rm e}^{ i \vec{k}\cdot\vec{x} } + \frac{ i k }{ 4\pi} H(k;k) j_0(kx)
    + {\bf P} \int \frac{ {\rm d}^3p }{ (2\pi)^3 }
        \frac{ H(p;k) j_0(px) }{ p^2 - k^2 }
\ ,
\label{BS_inf_F.eq}
\end{eqnarray}
where we neglect an irrelevant overall constant.

We assume
\begin{eqnarray}
      h(x;k)
  &\equiv& \int \frac{ {\rm d}^3p }{ (2\pi)^3 }
              H(p;k)
              {\rm e}^{ - i \vec{p} \cdot \vec{x} }
\cr
  &=& - ( \triangle + k^2 ) \phi_\infty (\vec{x};\vec{k})
   = 0  \qquad \mbox{ for $x > R$ }
\ ,
\label{Luscher_Cond_RFT.eq}
\end{eqnarray}
where $h(x;k)$ depends only on $x=|\vec{x}|$ and $k=|\vec{k}|$.
In order to simplify (\ref{BS_inf_F.eq}),
we use a formula :
\begin{equation}
    {\bf P} \int \frac{ {\rm d}^3p }{ (2\pi )^3 }
                            \frac{ F(p) }{ p^2 - k^2 }
   = \int {\rm d}^3 z \ f(z) \frac{ k }{ 4\pi } n_0(kz)
\ ,
\label{Formula_sim.eq}
\end{equation}
where $f(z)$ is an inverse Fourier transformation of $F(p)$
which is a function of $p=|\vec{p}|$.
From this
the third term in (\ref{BS_inf_F.eq}) can be written by
\begin{equation}
  E(x;k)
    \equiv
      {\bf P} \int \frac{ {\rm d}^3p }{ (2\pi)^3 }
                   \frac{ H(p;k) j_0(px) }{ p^2 - k^2 }
    = \int {\rm d}^3 z \ g(z;k,x) \frac{ k }{ 4\pi } n_0(kz)
\ ,
\label{EQrewrite_one.eq}
\end{equation}
where $g(z;k,x)$ is an inverse Fourier transformation of $H(p;k) j_0(px)$,
which is given by
\begin{eqnarray}
  g(z;k,x)
    &=& \int \frac{ {\rm d}^3p }{ (2\pi)^3 }
          H(p;k) j_0(px)
          {\rm e}^{ - i \vec{p}\cdot\vec{z} }
\cr
    &=& \int \frac{ {\rm d}\Omega_x }{ 4\pi }
        \int \frac{ {\rm d}^3p }{ (2\pi)^3 }
              H(p;k) {\rm e}^{   i \vec{p}\cdot \vec{x} }
                     {\rm e}^{ - i \vec{p}\cdot \vec{z} }
\cr
    &=& \int \frac{ {\rm d}\Omega_x }{ 4\pi }
        \int {\rm d}^3 y
              \, \delta^3 ( \vec{y} + \vec{x} - \vec{z} ) h(y;k)
\ ,
\label{EQrewrite_two.eq}
\end{eqnarray}
with the spherical coordinate $\Omega_x$ for $\vec{x}$.
By substituting (\ref{EQrewrite_two.eq}) into (\ref{EQrewrite_one.eq}),
we obtain
\begin{eqnarray}
  E(x;k)
   &=&  \int \frac{ {\rm d}\Omega_x }{ 4\pi }
        \int {\rm d}^3 y \ h(y;k)
                 \frac{ k }{ 4\pi } n_0(k|\vec{x}+\vec{y}|)
\cr
   &=&
        \int_{0}^\infty {\rm d}y \ ( 4\pi y^2 ) h(y;k)
        \int \frac{ {\rm d}\Omega_y }{ 4\pi }
                 \frac{ k }{ 4\pi } n_0(k|\vec{x}+\vec{y}|)
\cr
   &=&
        \int_{0}^\infty {\rm d}y \ ( 4\pi y^2 ) h(y;k)
        \frac{ k }{ 4\pi }
           \biggl[ \Theta(x-y) \cdot n_0(kx) j_0(ky)  \cr
   &&   \qquad\qquad\qquad\qquad\qquad
                   + \Theta(y-x) \cdot j_0(kx) n_0(ky) \biggr]
\ ,
\label{EQrewrite_three.eq}
\end{eqnarray}
where $\Theta(x-y)$ is the step function,
{\it i.e.} $\Theta(x-y)=1$ for $x-y\ge 0$ and $0$ for others.
Under the assumption (\ref{Luscher_Cond_RFT.eq}),
(\ref{EQrewrite_three.eq}) for $x>R$ can be written by
\begin{eqnarray}
        E(x;k)
    &=&
        \frac{ k }{ 4\pi } n_0(kx)
        \int_{0}^{R     } {\rm d}y \ ( 4\pi y^2 ) h(y;k) j_0(ky)
\cr
    &=&
        \frac{ k }{ 4\pi } n_0(kx)
        \int_{0}^{\infty} {\rm d}y \ ( 4\pi y^2 ) h(y;k) j_0(ky)
\cr
    &=&
        \frac{ k }{ 4\pi }n_0(kx)
        H(k;k)
\ .
\end{eqnarray}
Now we achieve the following simple expression of the BS function
in the infinite volume for $x>R$.
\begin{eqnarray}
       \phi_{\infty} (\vec{x};\vec{k})
   &=&
     {\rm e}^{ i \vec{k}\cdot\vec{x} }
        + \frac{ i k }{ 4\pi } H(k;k) j_0(kx)
        + \frac{   k }{ 4\pi } H(k;k) n_0(kx) 
\cr
   &=&
      \Bigl[ {\rm e}^{ i \vec{k}\cdot\vec{x} } - j_0(kx)  \Bigr]
        + {\rm e}^{ i\delta_0(k) } \cos \delta_0(k) j_0(kx)
        + {\rm e}^{ i\delta_0(k) } \sin \delta_0(k) n_0(kx)
\ ,
\label{BS_inf_FAS.eq}
\end{eqnarray}
where we use the relation (\ref{BS_def_infinitM2.eq})
in the final step.

The first term in (\ref{BS_inf_FAS.eq}) is written
in terms of $j_l(kx)$ as
\begin{equation}
   {\rm e}^{ i \vec{k} \cdot \vec{x} } - j_0(kx)
  = \sum_{l=1}^{\infty}
    \sum_{m=-l}^{l} (4\pi)
           i^l j_l(kx)
           Y_{lm}^* (\Omega_k)
           Y_{lm}   (\Omega_x)
\ ,
\label{plane_exp.eq}
\end{equation}
with the spherical coordinate
$\Omega_k$ for $\vec{k}$ and
$\Omega_x$ for $\vec{x}$.
Thus it does not contain the $S$-wave component
and it is a regular function for all $\vec{x}$.
We find that the ratio of the coefficients of $j_0(kx)$ and $n_0(kx)$
in (\ref{BS_inf_FAS.eq})
gives the $S$-wave scattering phase shift.
It should be noted that
$\phi_{\infty} (\vec{x};\vec{k})$
does not contain $n_l(kx)$ with $l\ge 1$.
This is attributed to the fact
that we neglect the scattering phase shifts $\delta_l(k)$ with $l \ge 1$
and regard $M(p;k)$ as a function of $p=|\vec{p}|$ and $k=|\vec{k}|$.

The BS function on a periodic box $L^3$
is defined by
\begin{equation}
    \phi (\vec{x};k) =
           \langle 0 | \pi_1(    \vec{x}/2 )
                       \pi_2(  - \vec{x}/2 )  | \pi_1 \pi_2 ; k \rangle
\ ,
\label{BS_fin_def.eq}
\end{equation}
where
$|\pi_1\pi_2;k\rangle$ is the energy eigenstate of two-pion system
on the periodic box with energy $E=2 \sqrt{m_\pi^2 + k^2}$.
Here we assume that the two-pion interaction range $R$ is smaller than
one half of the lattice extent, {\it i.e.} $R<L/2$,
so that the boundary condition does not distort
the shape of $h(x;k)$.

$\phi(\vec{x};k)$ should be written
in terms of $\phi_{\infty}(\vec{x};\vec{k})$
as follows.
\begin{equation}
   \phi (\vec{x};k) =
        \sum_{l=0}^{\infty}
        \sum_{m=-l}^{l}
           C_{lm}
           \sqrt{4\pi} Y_{lm}(\Omega_x)
           \phi^{lm}_{\infty} (x;k)
\ ,
\label{BS_rel.eq}
\end{equation}
where $C_{lm}$ are determined from the boundary condition
together with the allowed value of $k^2$,
and $\phi^{lm}_{\infty}(x;k)$ is the $l m$
component of $\phi_{\infty}(\vec{x};\vec{k})$
defined by
\begin{equation}
  \phi^{lm}_{\infty} (x;k)
  = \int \frac{ {\rm d}\Omega_x }{ 4\pi }
          \sqrt{4\pi} Y_{lm}^{*}(\Omega_x)
          \phi_{\infty}(\vec{x};\vec{k})
\ .
\end{equation}

$\phi(\vec{x};k)$ satisfies the the Helmholtz equation for $x>R$.
The general solution of the equation on a periodic box $L^3$
can be written by (\ref{Luscher_wave_one.eq})
with $G_{lm}(\vec{x};k)$ defined by (\ref{Glm_def.eq}).
Thus the equation (\ref{BS_rel.eq}) yields
\begin{eqnarray}
        \phi (\vec{x};k)
   &=&
        \sum_{l=0}^{\infty}
        \sum_{m=-l}^{l}
           C_{lm}
           \sqrt{4\pi} Y_{lm}(\Omega_x)
           \phi^{lm}_{\infty} (x;k)  \cr
   &=&  \sum_{l=0}^{\infty}
        \sum_{m=-l}^{l}
          v_{lm} G_{lm}(\vec{x};k)
\ ,
\label{BS_rel_two.eq}
\end{eqnarray}
for $x>R$.
As mentioned in Sec.~\ref{Sec:Luscher},
$G_{lm} (\vec{x};k)$ contains $n_{l'}(kx)$ with only $l'=l$.
$\phi_{\infty} (\vec{x};\vec{k})$
contains $n_l(kx)$ with only $l=0$
as known from (\ref{BS_inf_FAS.eq}) and (\ref{plane_exp.eq}).
Thus only $v_{00}$ is non-zero
and $G_{lm}(\vec{x};k)$ with $l\ge1$ do not contribute
in the second line of (\ref{BS_rel_two.eq}).

We use the expansion form of $G(\vec{x};k)$ given by (\ref{Exp_G.eq})
to determine the allowed value of $k^2$, $C_{lm}$ and $v_{00}$
in (\ref{BS_rel_two.eq}).
Comparing the $S$-wave component of both lines of (\ref{BS_rel_two.eq}),
we find
\begin{eqnarray}
&&  C_{00} \cdot {\rm e}^{ i \delta_0(k) } \sin\delta_0(k)
               = v_{00} \cdot \frac{k}{4\pi}
\label{BS_rel_four_1.eq} \\
&&  C_{00} \cdot {\rm e}^{ i \delta_0(k) } \cos\delta_0(k)
               = v_{00} \cdot g_{00}(k)
\ .
\label{BS_rel_four_2.eq}
\end{eqnarray}
Finally we obtain L\"uscher's formula (\ref{Lusher.eq})
by taking the ratio of (\ref{BS_rel_four_1.eq}) and (\ref{BS_rel_four_2.eq}).
\begin{equation}
  \frac{ 1 }{ \tan \delta_0( k )  }
  = \frac{ 4\pi }{ k } \cdot g_{00}(k)
\ .
\label{Luscher-BS.eq}
\end{equation}
The other components of (\ref{BS_rel_two.eq})
give only the informations for $C_{lm}$.

Let us make a comment on the condition (\ref{Luscher_Cond_RFT.eq}).
In quantum field theory
this condition is generally not exactly satisfied
and there can be a small tail in $h(x;k)$ for large $x$.
Thus we cannot rigorously define the two-pion interaction range.
Further
an exact estimation of the effects of the tail for the BS function
is not possible,
unless the off-shell two-pion scattering amplitude $H(p;k)$
for all $p$ for given $k$ is known.
In this appendix
we considered that the condition (\ref{Luscher_Cond_RFT.eq}) is
satisfied for some value of $R$,
with a tacit assumption that
the corrections from the interaction tails for the BS function are negligible.
%
%
\section{ Numerical calculation of periodic Green function }
\label{appe:zeta}
We rewrite $G(\vec{x};k)$ in terms of dimensionless values as
\begin{equation}
 ( 4\pi^2 L ) \cdot G(\vec{x};k)
  = \sum_{\vec{m}\in {\bf Z}^3}
    \frac{ {\rm e}^{ i \vec{m} \cdot \vec{X} } }{ m^2 - q_k^2 }
\ ,
\end{equation}
where
$q_k^2 = k^2 L^2 / (2\pi)^2$ ($\not\in {\bf Z}$)
and $\vec{X}=\vec{x} (2\pi)/L$.
The function $G(\vec{x};k)$ can be expanded
around the momentum $p^2 = n_p^2 \cdot (2\pi)^2/L^2$ ($n_p^2\in {\bf Z}$)
as
\begin{equation}
  ( 4 \pi^2 L ) \cdot G(\vec{x};k)
  = - \frac{1}{ q_k^2 - n_p^2 }
              \sum_{ m^2 = n_p^2 }
              {\rm e }^{ i \vec{m}\cdot\vec{X} }
  + \sum_{j=1}^{\infty} ( q_k^2 - n_p^2 )^{j-1} F( \vec{X}; j, n_p )
\ ,
\end{equation}
where
\begin{equation}
  F(\vec{X};j,n_p)
  = \sum_{ m^2 \not= n_p^2 }
    ( m^2 - n_p^2 )^{-j} \ {\rm e}^{ i \vec{m}\cdot \vec{X}}
\ .
\end{equation}
The function $F(\vec{X};j,n_p)$ depends on the position $\vec{X}$,
lattice geometry and the expansion point $n_p^2$.
But it is independent of the physical quantity,
such as quark masses and the strength of the two pion interaction.
In our case of the ground state of the two-pion, we set $n_p^2=0$.
We can use the same techniques as in Ref.~\cite{CP-PACS:phsh_full}
for the evaluation of the spherical zeta function.
$F(\vec{X};s,n_p)$ takes finite values for ${\rm Re}(s) > 3/2$.
The function at $s=j\in {\bf Z}\ge 1$
is defined by the analytic continuation from this region.

First we divide the summation in $F(\vec{X};s,n_p)$
into two parts as
\begin{equation}
  F(\vec{X};s,n_p)
  = \biggl[ \ \sum_{ m^2 < n_p^2 } + \sum_{ m^2 > n_p^2 } \ \biggr]
       \ ( m^2 - n_p^2 )^{-s}
       \ {\rm e}^{ i \vec{m}\cdot \vec{X}}
\ .
\end{equation}
The second part can be written by an integral form as
\begin{eqnarray}
&&   \sum_{ m^2 > n_p^2 }
                 \ ( m^2 - n_p^2 )^{-s}
                 \ {\rm e}^{ i \vec{m}\cdot \vec{X}}
\cr
&& = \biggl(
           \int_0^\infty {\rm d}t \sum_{ m^2 > n_p^2 } \
     \biggr)
           \frac{ t^{s-1} }{\Gamma(s)}
           {\rm e}^{ - t ( m^2 - n_p^2 ) }
           {\rm e}^{ i \vec{m}\cdot \vec{X}}
\cr
&& = \biggl(
          \int_1^\infty {\rm d}t \sum_{ m^2 >     n_p^2 }
       -  \int_0^1      {\rm d}t \sum_{ m^2 \le   n_p^2 }
       +  \int_0^1      {\rm d}t \sum_{ \vec{m}       }
    \biggr)
       \frac{ t^{s-1} }{\Gamma(s)}
       {\rm e}^{ - t ( m^2 - n_p^2 ) }
       {\rm e}^{ i \vec{m}\cdot \vec{X}}
\ .
\label{F_one.eq}
\end{eqnarray}
The first and second terms in (\ref{F_one.eq})
converge at $s=j \in {\bf Z} \ge 1$, which are given by
\begin{equation}
    - \frac{ 1 }{ j! }
         \sum_{ m^2 = n_p^2 }
         {\rm e}^{ i \vec{m}\cdot \vec{X}}
    - \sum_{ m^2 < n_p^2 }
         \frac{ {\rm e}^{ i \vec{m}\cdot \vec{X}} }{ ( m^2 - n_p^2 )^j }
    + \sum_{ r = 1}^{ j }
         \frac{ 1 }{ ( j - r )!}
         \sum_{ m^2 \not= n_p^2 }
           \frac{ {\rm e}^{ - ( m^2 - n_p^2 ) } }{ ( m^2 - n_p^2 )^r }
           \ {\rm e}^{ i \vec{m}\cdot \vec{X}}
\ ,
\end{equation}
at $s=j \in {\bf Z} \ge 1$.
The third term in (\ref{F_one.eq}) is
rewritten by Poisson's summation formula :
\begin{equation}
    \sum_{ \vec{m} \in {\bf Z}^3 }  f(\vec{m})
  = \sum_{ \vec{m} \in {\bf Z}^3 }
       \int {\rm d}^3y \ f(\vec{y})
           {\rm e}^{ i 2 \pi \vec{m} \cdot \vec{y} }
\ ,
\label{F_three.eq}
\end{equation}
and integration over $\vec{y}$ yields,
\begin{eqnarray}
      \int_0^1 {\rm d}t
      \sum_{ \vec{m} \in {\bf Z}^3 }
      \frac{ t^{s-1} }{ \Gamma (s)}
      {\rm e}^{ - t ( m^2 - n_p^2 ) } \ {\rm e}^{ i \vec{m}\cdot \vec{X}}
  =
      \int_0^1 {\rm d}t
      \frac{ t^{s-1} }{ \Gamma (s)}
      \biggl( \frac{\pi}{t} \biggr)^{3/2}
      \ {\rm e}^{ t n_p^2 }
      \sum_{ \vec{m} \in {\bf Z}^3 }
         {\rm e}^{ - ( \vec{X} + 2\pi \vec{m} )^2 / ( 4 t )  }
\ .
\label{F_two.eq}
\end{eqnarray}
The final expression in (\ref{F_two.eq})
converges at $s=j \in {\bf Z} \ge 1$
for $\vec{X}/(2\pi) \not\in {\bf Z}^3$.
We do not need the values
at $\vec{X}/(2\pi) = \vec{n}\in {\bf Z}^3$,
because these correspond to $\vec{x} = \vec{n} L$ and
these positions are within the two-pion interaction range.
Finally, gathering all terms and setting
$s=j \in {\bf Z} \ge 1$ in (\ref{F_two.eq}),
we obtain
\begin{eqnarray}
&&
   F( \vec{X} ; j , n_p ) =
    - \frac{ 1 }{ j! }
      \sum_{ m^2 = n_p^2 }
      {\rm e}^{ i \vec{m}\cdot \vec{X}}
    + \sum_{ r = 1}^{ j }
        \frac{ 1 }{ ( j - r )! }
        \sum_{ m^2 \not= n_p^2 }
        \frac{  {\rm e}^{ - ( m^2 - n_p^2 ) } }{ ( m^2 - n_p^2 )^r }
        \ {\rm e}^{ i \vec{m}\cdot\vec{X}}
\cr
&& \qquad\qquad \qquad\qquad
   +
      \frac{ \pi^{3/2} }{ ( j - 1 )! }
      \int_0^1 {\rm d}t \ t^{j-5/2} \ {\rm e}^{ t n_p^2 }
      \sum_{ \vec{m} }
         {\rm e}^{ - ( \vec{X} + 2\pi \vec{m} )^2 / (4t) }
\ \ \ .
\end{eqnarray}
%
%
\newpage
%
%

%
%
\newpage
%
%
%
%
%
%
%
%
%
%
%
%
%
%
%
%
%
%
%
%
%
%
%
%
%
\begin{table}[t]
\begin{center}
\begin{tabular}{ll lllll}
\hline
\hline
$L^3=24^3$ \\
\hline
\multicolumn{2}{l}{ $m_\pi \ ({\rm GeV})$ }
& $ 0.85773(27) $
& $ 0.76675(28) $
& $ 0.66649(30) $
& $ 0.59205(31) $
& $ 0.52290(33) $ \\
\multicolumn{2}{l}{ $m_\pi^2 \ ({\rm GeV}^2)$ }
& $ 0.73570(47) $
& $ 0.58790(43) $
& $ 0.44421(40) $
& $ 0.35052(37) $
& $ 0.27342(35) $ \\
\multicolumn{2}{l}{ $x_m$ }
& $ 9.8 $
& $ 9.8 $
& $ 9.8 $
& $ 9.4 $
& $ 9.0 $ \\
\multicolumn{2}{l}{ $\Delta W \ (\times 10^{-3}\ {\rm GeV})$ } \\
& from $T$
& $ 1.748(64) $
& $ 2.011(73) $
& $ 2.337(86) $
& $ 2.60 (10) $
& $ 2.85 (11) $ \\
& from $V$
& $ 1.817(25) $
& $ 2.049(29) $
& $ 2.336(35) $
& $ 2.559(37) $
& $ 2.804(40) $ \\
& from $\phi$
& $ 1.832(43) $
& $ 2.071(49) $
& $ 2.346(58) $
& $ 2.550(65) $
& $ 2.784(66) $ \\
\multicolumn{2}{l}{ $k^2 \ (\times 10^{-3}\ {\rm GeV}^2)$ } \\
& from $T$
& $ 1.500(55) $
& $ 1.543(56) $
& $ 1.559(58) $
& $ 1.540(59) $
& $ 1.492(60) $ \\
& from $V$
& $ 1.559(22) $
& $ 1.572(22) $
& $ 1.558(23) $
& $ 1.517(22) $
& $ 1.468(21) $ \\
& from $\phi$
& $ 1.572(36) $
& $ 1.589(38) $
& $ 1.565(39) $
& $ 1.512(39) $
& $ 1.457(35) $ \\
\multicolumn{2}{l}{ $a_0/m_\pi \ (1/{\rm GeV}^2)$ } \\
& from $T$
& $ -0.969(32) $
& $ -1.112(36) $
& $ -1.291(42) $
& $ -1.439(49) $
& $ -1.585(57) $ \\
& from $V$
& $ -1.003(12) $
& $ -1.131(14) $
& $ -1.291(17) $
& $ -1.420(18) $
& $ -1.562(20) $ \\
& from $\phi$
& $ -1.010(21) $
& $ -1.141(24) $
& $ -1.296(29) $
& $ -1.415(32) $
& $ -1.552(33) $ \\
\hline
\hline
\end{tabular}
\caption{
Results on $24^3$ lattice from
two-pion time correlator (``from $T$''),
constant fit of $V(\vec{x};k)$ (``from $V$'') and
fitting wave function
with $G(\vec{x};k)$ (``from $\phi$'').
The fitting range of $V(\vec{x};k)$ and $\phi(\vec{x};k)$ are quoted as $x_m$.
}
\label{SCL-POT-ZAN-RXPA.2480.KX.table}
\end{center}
%
\newpage
\end{table}
%
%
\begin{table}[t]
\begin{center}
\begin{tabular}{ll lllll}
\hline
\hline
$L^3=20^3$ \\
\hline
\multicolumn{2}{l}{ $m_\pi \ ({\rm GeV})$ }
& $ 0.85754(25) $
& $ 0.76653(26) $
& $ 0.66628(28) $
& $ 0.59188(30) $
& $ 0.52283(32) $ \\
\multicolumn{2}{l}{ $m_\pi^2 \ ({\rm GeV}^2)$ }
& $ 0.73537(42) $
& $ 0.58757(40) $
& $ 0.44393(38) $
& $ 0.35033(36) $
& $ 0.27335(34) $ \\
\multicolumn{2}{l}{ $x_m$ }
&  &  &  & $ 9.0 $  & $ 8.2 $ \\
\multicolumn{2}{l}{ $\Delta W \ (\times 10^{-3}\ {\rm GeV})$ } \\
& from $T$
& $ 3.219(88) $
& $ 3.66 (10) $
& $ 4.20 (11) $
& $ 4.63 (13) $
& $ 5.03 (15) $ \\
& from $V$
&  &  &  & $ 4.457(63) $  & $ 4.928(68) $ \\
& from $\phi$
&  &  &  & $ 4.62(12)  $  & $ 5.05(12)  $ \\
\multicolumn{2}{l}{ $k^2 \ (\times 10^{-3}\ {\rm GeV}^2)$ } \\
& from $T$
& $ 2.763(76) $
& $ 2.810(77) $
& $ 2.805(76) $
& $ 2.744(75) $
& $ 2.635(78) $ \\
& from $V$
&  &  &  & $ 2.643(38) $  & $ 2.583(36) $ \\
& from $\phi$
&  &  &  & $ 2.741(72) $  & $ 2.646(64) $ \\
\multicolumn{2}{l}{ $a_0/m_\pi \ (1/{\rm GeV}^2)$ } \\
& from $T$
& $ -1.003(24) $
& $ -1.139(27) $
& $ -1.310(31) $
& $ -1.447(35) $
& $ -1.584(41) $ \\
& from $V$
&  &  &  & $ -1.401(17) $  & $ -1.556(19) $ \\
& from $\phi$
&  &  &  & $ -1.446(33) $  & $ -1.589(34) $ \\
\hline
\hline
\end{tabular}
\caption{
Results on $20^3$ lattice from
two-pion time correlator (``from $T$''),
constant fit of $V(\vec{x};k)$ (``from $V$'') and
fitting wave function
with $G(\vec{x};k)$ (``from $\phi$'').
The fitting range of $V(\vec{x};k)$ and $\phi(\vec{x};k)$ are quoted as $x_m$.
}
\label{SCL-POT-ZAN-RXPA.2080.KX.table}
\end{center}
%
\newpage
\end{table}
%
%
\begin{table}[t]
\begin{center}
\begin{tabular}{ll lllll}
\hline
\hline
$L^3=16^3$ \\
\hline
\multicolumn{2}{l}{ $m_\pi \ ({\rm GeV})$ }
& $ 0.85759(34) $
& $ 0.76672(37) $
& $ 0.66660(41) $
& $ 0.59225(45) $
& $ 0.52318(50) $ \\
\multicolumn{2}{l}{ $m_\pi^2 \ ({\rm GeV}^2)$ }
& $ 0.73546(58) $
& $ 0.58786(57) $
& $ 0.44435(55) $
& $ 0.35076(53) $
& $ 0.27371(52) $ \\
\multicolumn{2}{l}{ $x_m$ }
&  &  &  &  & $ 7.8 $ \\
\multicolumn{2}{l}{ $\Delta W \ (\times 10^{-3}\ {\rm GeV})$ } \\
& from $T$
& $ 6.51(17) $
& $ 7.28(20) $
& $ 8.20(23) $
& $ 8.93(26) $
& $ 9.62(32) $ \\
& from $V$
& & & & & $ 9.64(21) $ \\
& from $\phi$
& & & & & $ 9.71(41) $ \\
\multicolumn{2}{l}{ $k^2 \ (\times 10^{-3}\ {\rm GeV}^2)$ }  \\
& from $T$
& $ 5.59(15) $
& $ 5.59(15) $
& $ 5.48(15) $
& $ 5.31(16) $
& $ 5.06(17) $ \\
& from $V$
& & & & & $ 5.07(11)  $ \\
& from $\phi$
& & & & & $ 5.10(22) $ \\
\multicolumn{2}{l}{ $a_0/m_\pi \ (1/{\rm GeV}^2)$ }  \\
& from $T$
& $ -1.002(22) $
& $ -1.122(26) $
& $ -1.270(30) $
& $ -1.393(35) $
& $ -1.517(43) $ \\
& from $V$
& & & & & $ -1.519(28) $ \\
& from $\phi$
& & & & & $ -1.528(55) $ \\
\hline
\hline
\end{tabular}
\caption{
Results on $16^3$ lattice from
two-pion time correlator (``from $T$''),
constant fit of $V(\vec{x};k)$ (``from $V$'') and
fitting wave function
with $G(\vec{x};k)$ (``from $\phi$'').
The fitting range of $V(\vec{x};k)$ and $\phi(\vec{x};k)$ are quoted as $x_m$.
}
\label{SCL-POT-ZAN-RXPA.1680.KX.table}
\end{center}
\end{table}
%
%
\end{document}